\documentclass[%
 preprint, 
 amsmath,amssymb,
 aps, physrev,
]{revtex4-2}
\usepackage{ragged2e}
\usepackage{graphicx}
\usepackage{dcolumn}
\usepackage{bm}
\usepackage{tikz}
\usepackage{array}
\usepackage{siunitx} 
\usepackage{booktabs} 
\usetikzlibrary{patterns}

\DeclareRobustCommand{\legendsquare}[1]{%
  \tikz[baseline=(a.south)]{\node[#1, inner sep=.4ex, outer sep=0] (a) {};}%
}
\newcommand{\blackline}{\raisebox{2pt}{\tikz{\draw[-,black!40!black,solid,line width = 0.9pt](0,0) -- (5mm,0);}}}

\newcommand{\bluecross}{\raisebox{1pt}{\tikz{\draw[blue!40!blue] (-0.75mm,-0.75mm) -- (0.75mm,0.75mm) (-0.75mm,0.75mm) -- (0.75mm,-0.75mm);}}}

\newcommand{\blackdashed}{\raisebox{2pt}{\tikz{\draw[-,black,dashed,line width = 0.9pt](0,0) -- (5mm,0);}}}

\usepackage{hyperref}

\usepackage[process=all]{pstool}
\makeatletter
\providecommand\HyPL@Entry[1]{}
\AddToHook{env/document/begin}{%
\@ifpackageloaded{preview}{
\ifPreview
 \let\Hy@FirstPageHook\relax
 \let\Hy@EveryPageAnchor\relax
\fi}{}}
\makeatother
\begin{document}

\preprint{APS/123-QED}

\title{\textbf{Three-dimensional variational data assimilation of separated flows using time-averaged experimental data} 
}%

\author{Uttam Cadambi Padmanaban}
 \email{ucp1n22@soton.ac.uk}
\author{Bharathram Ganapathisubramani}%
 \email{g.bharath@soton.ac.uk}
\author{Sean Symon}%
 \email{sean.symon@soton.ac.uk}
\affiliation{%
Department of Aeronautics and Astronautics, University of Southampton, \\United Kingdom
}%

\date{\today}

\begin{abstract}
We present a novel framework for assimilating planar PIV experimental data using a variational approach to enhance the predictions of the Spalart–Allmaras RANS turbulence model. Our method applies three-dimensional constraints to the assimilation of mean velocity data, incorporating a corrective forcing term in the momentum equations. The advantages of this approach are highlighted through a direct comparison with traditional two-dimensional assimilation using the same experimental dataset. We demonstrate its efficacy by assimilating the deep stall flow over a NACA0012 airfoil at a $15^\circ$ angle of attack and a chord-based Reynolds number of $Re_c \approx 7.5 \times 10^4$. We find that in two-dimensional assimilation, the corrective forcing term compensates not only for physical modeling errors but also for the lack of divergence in the experimental data. This conflation makes it difficult to isolate the effects of measurement inconsistencies from deficiencies in the turbulence model. In contrast, three-dimensional assimilation allows the corrective forcing term to primarily address experimental setup errors while enabling the turbulence model to more accurately capture the flow physics. We establish the superiority of three-dimensional assimilation by demonstrating improved agreement in reconstructed quantities, including pressure, lift force, and Reynolds shear stress.

\end{abstract}
\maketitle
\newpage


\section{Introduction}
\label{section:introduction}

Data assimilation (DA) first emerged in meteorology \cite{le1986variational}, where it was used to improve weather predictions by incorporating observational data into numerical models. DA helps reduce observational noise and uncertainties, expand the field of view of available data, and refine modeling assumptions by calibrating model coefficients and boundary conditions. These advantages have led to its increasing use in fluid mechanics, particularly for studying turbulent flows, as demonstrated by recent studies \cite{foures2014data, symon2019tale, franceschini2020mean, thompson2024effect, mons2024data}. The complexity of turbulence arises from its wide range of length and time scales and its inherently unsteady behavior. Resolving all these scales using direct numerical simulations (DNS) or large eddy simulations (LES) is computationally intractable for high Reynolds number flows of practical interest \cite{pope2001turbulent}. A more computationally efficient alternative is the use of Reynolds-Averaged Navier–Stokes (RANS) turbulence models, which rely on modeling assumptions to predict an averaged flow state. However, their accuracy is questionable in cases involving strong pressure gradients and flow separation, as they are typically calibrated against only a limited set of canonical flow problems. The application of DA to fluid mechanics problems is particularly useful in improving RANS turbulence models that are able to capture the effects of turbulence a lot more accurately. 

The widespread adoption of DA in fluid mechanics has also been driven by advances in field imaging techniques, such as planar particle image velocimetry (PIV), which provides access to two-component mean velocity data \cite{adrian1991particle}. PIV is often used as an alternative to DNS or LES, particularly for studying high Reynolds number flows of practical interest. However, like any experimental method, PIV is subject to noise and uncertainty. It also comes with additional limitations, including low resolution, a limited field of view, high setup costs (especially due to the laser), large data storage requirements, and long processing times. An important problem that is always overlooked is the lack of the third (out-of-plane) component of velocity which is not imaged using planar PIV, and is non-negligible for separated flows even at low Reynolds number \cite{barkley1996three,gupta2023two}. Assimilating two-component mean velocity data from planar PIV without enforcing three-dimensional (3D) constraints leads to an inaccurate representation of the flow dynamics, as the planar PIV data do not satisfy the 2D continuity equation since the actual flow is three-dimensional. 

 Among the different types of DA, we use the variational method applied to steady-state cases in this study. Variational DA optimizes a control variable of choice by minimizing the discrepancy between a model output (such as RANS) and reference data (either from experiments or high fidelity simulation). Gradient-based optimization methods are used to minimize the objective function that quantitatively describes the deviation of model output from reference data. The time-averaged continuity and momentum equations are typically applied as constraints. Adjoint methods are used as an efficient way of computing gradients \cite{peter2010numerical, kenway2019effective}. The application of 3D DA has been limited due to a variety of reasons. Reference \cite{foures2014data} performed variational DA of a flow past a circular cylinder at $Re = 150$ using two-component, mean velocity fields from DNS as reference data. A forcing term in the time-averaged momentum equations was optimized without the application of a turbulence model. The two-dimensional (2D) continuity equation was enforced as a constraint, which is consistent with the 2D nature of the flow \cite{barkley1996three}. The use of experimental data using a similar methodology for assimilating a higher Reynolds number ($Re_c = 13,500$) flow past an idealized airfoil was demonstrated by Ref. \cite{symon2017data}. The justification for enforcing 2D continuity was offered by comparing the optimized forcing with the forcing obtained from the experiment. A good agreement between the two was used to argue that the field was indeed two-dimensional.

Reference \cite{franceschini2020mean} extended the work of Ref. \cite{foures2014data} by including the Spalart-Allmaras (SA) turbulence model for assimilating the mean velocity field into the RANS equations. This improved the efficiency of the assimilation for high Reynolds number flows as the baseline solution was already a close approximation of the reference mean velocity field. The authors optimized a control variable in the time-averaged momentum equations (similar to Refs. \cite{foures2014data, symon2017data}) of a flow past a backward-facing step (BFS) at Reynolds number $Re = 28275$ using DNS reference data. The study utilized a span-averaged and time-averaged velocity field permitting the enforcement of 2D continuity. Additionally, the reference data were obtained from a strictly divergence-free space of mean velocity fields. This meant that the obtained forcing field was only compensating for the lack of physics that the turbulence model was unable to capture and not any deficiency in the quality of reference data. Recently, Ref. \cite{mons2024data} used a corrective source term in the turbulence transport equation of the SA model for mean velocity assimilation of a flow past a NACA0012 airfoil. The near-stall flow was at an angle of attack $\alpha = 10^\circ$ and in the Reynolds number range $4.3 \times 10^4 \leq Re \leq 6.4 \times 10^4$. The reference mean velocity field was obtained from time-averaged PIV. DA was able to fill in the gaps that resulted from shadows and reflections during the data acquisition stage of PIV. The effects of three-dimensionality on reconstructed flow quantities such as Reynolds stresses, nonetheless, were not investigated.

The studies discussed so far made use of the continuous adjoint approach, which involves linearizing the PDE followed by its discretization. This way, the primal solver used to solve the Navier-Stokes (NS) equations can be re-used for solving the adjoint equations. The discrete adjoint approach, as demonstrated by Ref. \cite{kenway2019effective} constructs the adjoint equations following its discretization, driving the accuracy of gradient calculation to machine precision. References \cite{brenner2022efficient, brenner2024variational} used a discrete adjoint approach with reference data obtained from LES, which is free from noise, and, more importantly, allows the application of 2D constraints. Reference \cite{cato2023comparison} compared various control variables for 2D mean velocity assimilation, reporting good agreement with reference data and satisfactory Reynolds stress reconstruction. 

The need for high-accuracy gradients has led to the emergence of tools that have a discrete adjoint solver implemented in them. DAFoam \cite{he2020dafoam, he2018aerodynamic} is one such tool that incorporates a discrete adjoint solver along with primal solvers from OpenFOAM \cite{weller1998tensorial}. Since the code is open source, easy modifications and extensions to the solvers and turbulence models are possible. One such useful extension is the development of a projection and smoothing operation (see Refs. \cite{symon2017data, symon2019tale}) that allows the use of experimental reference data \cite{thompson2024effect}. Reference \cite{thompson2024effect} used DAFoam and compared the effects of input data resolution on the assimilation accuracy by constructing synthetic data sets from DNS. Two-dimensional constraints could be correctly applied here due to the nature of the reference data used. Three-dimensional constraints have been applied using DAFoam to assimilate a 3D flow past a wall-mounted cube with a single plane of measurements generated using a hybrid RANS/LES model, demonstrating its feasibility in 3D assimilation, albeit with all three velocity components \cite{cadambi2024towards}. The operations defined in OpenFOAM and the ease of running applications on high-performance computing (HPC) clusters make DAFoam an attractive choice for performing 3D DA. 

The complexity of performing 3D DA extends beyond traditional variational methods. A more recent technique, physics-informed neural networks (PINNs) \cite{raissi2019physics}, integrates the governing equations of fluid flow into the training of neural networks. Unlike traditional neural networks, PINNs enforce these equations as soft constraints by minimizing the residuals of the governing equations, ensuring that the model adheres to physical laws while learning the solution. PINNs have been applied to fluid mechanics problems ranging from simple problems to demonstrate the method \cite{sliwinski2023mean} to problems that have integrated turbulence models into the network \cite{patel2024turbulence}. Most of these studies use DNS or LES reference data to train the model. Some studies have explored the use of experimental data for jets \cite{von2022mean} magnetic resonance imaging to correct noisy data and displacement artifacts \cite{villie2025physics}, and a more complicated case of stalled flow past an airfoil at high Reynolds number \cite{christian2025review}. The challenge of enforcing a 2D constraint for 3D data may not be as severe in PINNs since the constraints are not enforced as strongly as in the case of the variational method. 

Most DA studies reviewed so far have focused on reconstructing flow quantities for canonical 2D problems. Despite the large-scale benefits of variational DA applied to steady-state cases, it is a technique that is undergoing development with limitations in assimilating 3D flows with limited observational data. This is reflected in the range of Reynolds numbers investigated (with the exception of Ref. \cite{mons2024data}) and the primary use of high-fidelity reference data such as DNS/LES (with the exceptions of Refs. \cite{symon2017data, mons2024data}). The majority of corrections applied in these studies have targeted the turbulence model \cite{franceschini2020mean, mons2016reconstruction, thompson2024effect, cadambi2024towards}, which has shown success for flows with moderate separation. When using high-fidelity simulations as reference data, the continuity constraint is strongly enforced, eliminating divergence errors. For weakly separated flows, span-averaging ensures a quasi-two-dimensional behavior, making the application of 2D continuity constraints reasonable during assimilation.

However, for high Reynolds number flows with strong separation and complex dynamics, access to high-fidelity data that strictly satisfies 2D continuity in a span-averaged sense is limited. In such cases, planar PIV reference data are often used. Experimental PIV data not only contains noise and uncertainties but also does not inherently satisfy 2D continuity. Performing 2D DA on such data while enforcing 2D continuity constraints leads to misinterpretation of the corrective/control variables, as briefly noted in Ref. \cite{symon2017data}. This misinterpretation is problematic for developing generalizable models (as in Ref. \cite{volpiani2021machine}) and for deriving accurate second-order statistics, such as Reynolds shear stress, from assimilated mean velocity fields. When dealing with flows where the out-of-plane velocity component is significant, a 3D assimilation approach is necessary to accurately recover flow physics and improve the interpretation of the control variables.  We present a novel approach for assimilating planar PIV data of a flow past a NACA0012 airfoil in deep stall at Reynolds number $Re_c \approx 7.5 \times 10^4$ \cite{carter2023low} using variational data assimilation (DA) with a discrete adjoint method. This method introduces a corrective forcing term in the momentum equations as a control variable while enforcing 3D constraints. We compare its performance against conventional variational DA with 2D constraints and re-evaluate the role of the corrective forcing and turbulence model. 

The rest of the paper is organized as follows. The governing equations of fluid flow and the variational DA methodology are described in Sec. \ref{section:methods}, where we provide a comprehensive explanation of the assimilation framework. This section also describes the computational setup of 2D DA and introduces our novel method of 3D DA that is used in the study. In Sec. \ref{section:flow_configuration_and_reference_data}, we present a NACA0012 airfoil in deep stall and discuss both the synthetic and experimental reference data sets. The results from baseline computations are reported in Sec. \ref{section:baseline_computation}, followed by the assimilation of synthetic data in Sec. \ref{section:assimilation_of_synthetic_data}. We compare the assimilation of experimental data using 2D constraints in Sec. \ref{section:2DVar_assimilation_of_experimental_data} and 3D constraints in Sec. \ref{section:3DVar_assimilation_of_experimental_data}. The accuracy and impact on the reconstructed flow variables are analyzed in Sec. \ref{section:reconstructed_quants}, with concluding remarks summarized in Sec. \ref{section:conclusion}.
 
\section{Turbulence modeling and data assimilation}
\label{section:methods}
In this section, a discussion of the fluid model is provided along with the DA method used. Section \ref{subsection:RANS} presents the governing equations of fluid flow. Section \ref{subsection:data_assimilation} provides a detailed description of the variational formulation and proceeds to motivate the choice of the control variable, where a forcing term is added to the momentum equations. In Sec. \ref{subsection:data_assimilation_setup}, we describe the methodology of 3D DA and distinguish it from a traditional 2D DA procedure.

\subsection{Reynolds-Averaged Navier--Stokes}
\label{subsection:RANS}
We are interested in the equations governing the ensemble-averaged flow. To this end, the ensemble-averaged (RANS) equations for an incompressible fluid are given by,
\begin{gather}
\frac{\partial {U_i}}{\partial x_i} = 0, \\
\label{equation:RANS}
U_j \frac{\partial U_i}{\partial x_j} = -\frac{1}{\rho} \frac{\partial P^*}{\partial x_i} + \frac{\partial}{\partial x_j} \left( \nu \frac{\partial U_i}{\partial x_j} \right) - \frac{\partial  \tau_{ij}}{\partial x_j},
\end{gather}
where $U_i$ and $P^*$ are the mean velocity components and pressure, respectively, $\rho$ is the density of the fluid, $\nu$ is the kinematic viscosity, $\tau_{ij}$ is the Reynolds stress tensor, and $x_i$ represents the spatial coordinates. The Reynolds stress tensor $\tau_{ij} = \overline{u_i' u_j'}$ is defined as the averaged outer product of the fluctuating velocity components, presenting the well-known problem of closure. Here, the notation $<\cdot>'$ denotes fluctuations, representing the deviation of the instantaneous velocity from its ensemble average, expressed as
\begin{align}
u_i' = u_i - U_i,
\end{align}
where $u_i$ is the instantaneous velocity, and $U_i$ is the ensemble average. The overbar $<\overline{\cdot}>$ denotes Reynolds averaging.

\subsection{Data assimilation methdology}
\label{subsection:data_assimilation}
We expect that the solution to the RANS equations, as defined in Sec. \ref{subsection:RANS}, will deviate from experimental observations, necessitating the use of DA. To apply variational DA, we require the definition of a control variable and an objective function. Following the approach of Ref. \cite{foures2014data}, we use a forcing term added to the momentum equations as the control variable. The modified momentum equations, incorporating this forcing term, are given by
\begin{gather}
\label{equation:mom_correction}
U_j \frac{\partial U_i}{\partial x_j} + \frac{1}{\rho} \frac{\partial P^*}{\partial x_i} - \frac{\partial}{\partial x_j} \left( \nu \frac{\partial U_i}{\partial x_j} \right) - f_{R_i} - f_{c_i} = 0,
\end{gather}
where $f_{c_i}$ is the control variable applied as a corrective forcing in addition to $f_{R_i} = - \partial \tau_{ij}/\partial x_j$,  the modeled Reynolds forcing. Using the Helmholtz decomposition, the corrective forcing term $f_{c_i}$ can be decomposed into the sum of a solenoidal part $f_{s_i}$ and an irrotational part given by,
\begin{gather}
    \label{equation:helmholtz}
    f_{c_i} = f_{s_i} + \frac{1}{\rho}\frac{\partial \phi}{\partial x_i}. 
\end{gather}
The irrotational part is lumped in with the pressure $p^* \approx p - \phi$. The consequence of such a decomposition is that the recovery of the irrotational term is not possible unless pressure data are supplied.

The objective function is the discrepancy between high-fidelity/experimental and RANS simulation mean velocity fields,
\begin{align}
\label{equation:mom_source_obj_func}
    f(\mathbf{u}, \mathbf{f}_c) = \frac{1}{2} \|\mathcal{Q}(\mathbf{u}, \mathbf{f}_c) - \Tilde{\mathbf{Q}}\|_Q^,
\end{align}
where $\Tilde{\mathbf{Q}}$ is a set of high-fidelity data/experimental measurements and bold-face symbols denote vectors. Operator $\mathcal{Q} (.)$ extracts the computational data in such a way that $\mathcal{Q}(\mathbf{u}) \in Q$ is a projection of the computational mean velocity to the measurement space $Q$. \(\|\cdot\|_Q\) is the generic norm in the measurement space. 

Variational DA is now formulated as an optimization problem where the goal is to minimize an objective function subject to some constraints. This is mathematically written as
    \begin{align}
    \label{equation:constopti}
\min_{\mathbf{w} \in \mathbb{R}^{n_w} , \mathbf{f}_c \in \mathbb{R}^{n_{f_c}}} \quad & f(\mathbf{w}, \mathbf{f}_c),\\
    \label{equation:constopti1}
\textrm{s.t.} \quad & R(\mathbf{w}, \mathbf{f}_c) = 0, \\
    \label{equation:constopti2}
\quad & \mathbf{f}_{c_L} \leq \mathbf{f}_c \leq \mathbf{f}_{c_U}, 
\end{align}
where $n_{f_c}$ is the size of the control variable, $n_w$ is the size of the state vector, $R$ is the governing equations that serve as constraints and $\mathbf{f}_{c_L}$ and $\mathbf{f}_{c_U}$ denote the lower and upper bounds, respectively, for the control variable.  For our test case, $R$ represents the residual of the NS equations. Equations \ref{equation:constopti} - \ref{equation:constopti2} form a non-linear, constrained minimization problem with equality and bound constraints and can be solved using gradient-based techniques. 

Gradient-based optimization techniques require the total derivative of the objective function with respect to the control variable (hereafter referred to as sensitivity). An efficient way to compute the sensitivity is by employing an adjoint method, which ensures that the computational cost remains independent of the number of control variables \citep{giannakoglou2008adjoint}. We use the discrete adjoint method in this study for computing the sensitivities. If $f$ and $R$ are a univariate representation of the objective function and residual of the NS equations, respectively, the sensitivity can be computed using
\begin{align}
\label{equation:discadjfinal}
   \frac{df}{d \mathbf{f}_c} =\frac{\partial f}{\partial \mathbf{f}_c} - \psi^T \frac{\partial \mathbf{R}}{\partial \mathbf{f}_c},
\end{align}
where $\psi^T$ is the transpose of the adjoint vector. The detailed derivation can be found in Ref. \citep{kenway2019effective}. DAFoam is used to obtain the sensitivity. DAFoam's source code is enriched with AD-forward (ADF) and AD-reverse (ADR) implementations using CoDiPack \citep{SaAlGauTOMS2019}, enabling machine-precision gradient accuracy \cite{kenway2019effective}. 
Once the sensitivity is obtained, we use sequential quadratic programming (SQP) \citep{wilson1963simplicial} to perform the optimization. The SQP method solves an optimization problem by obtaining search directions from a sequence of quadratic programming (QP) sub-problems with linearized constraints. The sub-problem defines a quadratic model of a certain Lagrangian with linear equality and inequality constraints. A Newton method is used along with a line search filter that minimizes a penalty function to obtain search directions for the QP sub-problems. The initial set of iterates is then updated based on these search directions. We use SNOPT (Sparse Non-linear Optimizer)[Ref. \citep{gill2005snopt}] which exploits the sparsity in the constraint Jacobian and has implementations that can tackle the issues mentioned earlier.

We implement a cell-volume weighted averaging operator $\mathcal{Q}$ that was used in Ref. \cite{symon2017data} to project the computational mean velocity data onto the experimental grid. While this may be trivially accomplished using standard interpolation methods for high-fidelity reference data, the same cannot be employed for noisy experimental data. The cell-weighted averaging procedure used in this study does not exploit cell-cell intersections for computing the volume-averaged mean velocity field as is done in Ref. \cite{thompson2024effect}. Once the mean velocity data from the computational grid is projected onto the experimental grid, the objective function is calculated. Since the adjoint solution is forced on the computational grid, a smoothing operator $\mathcal{\hat{Q}}$ is used to transfer the discrepancy back to the computational grid. The implementation of this projection and smoothing operation is essential to ease the cost of calculating the discrepancy field for 3D problems. A custom objective function is also implemented that can utilize the projection and smoothing operations. The details of the implementation are presented in Appendix \ref{appendix:projection_smoothing}. 

\subsection{Data assimilation setup}
\label{subsection:data_assimilation_setup}
Two-dimensional DA (hereafter referred to as 2DVar) applies the governing equations in Eq. \ref{equation:RANS} and the constraints in Eq. \ref{equation:constopti1} in only two spatial dimensions: streamwise and wall-normal. This approach has been the standard in all studies utilizing variational DA so far. However, for a truly 3D flow, enforcing 2D constraints leads to a misrepresentation of both the control variable and the role of the turbulence model. In this case, the control variable does not only compensate for the limitations of the turbulence model but also for the missing 3D effects, leading to an inaccurate correction. To address this, we propose a novel approach - 3D DA (hereafter referred to as 3DVar)—where the governing equations and constraints incorporate all three spatial dimensions: streamwise, wall-normal, and spanwise. This ensures that the 3D nature of the flow is properly accounted for, leading to a more physically meaningful interpretation of the control variable. We present the setups of both 2DVar and 3DVar and emphasize that, regardless of the chosen setup, the reference data always consist of a two-component mean velocity field.


\begin{figure}[h!]
\centering
\psfragfig[width=0.7\textwidth]{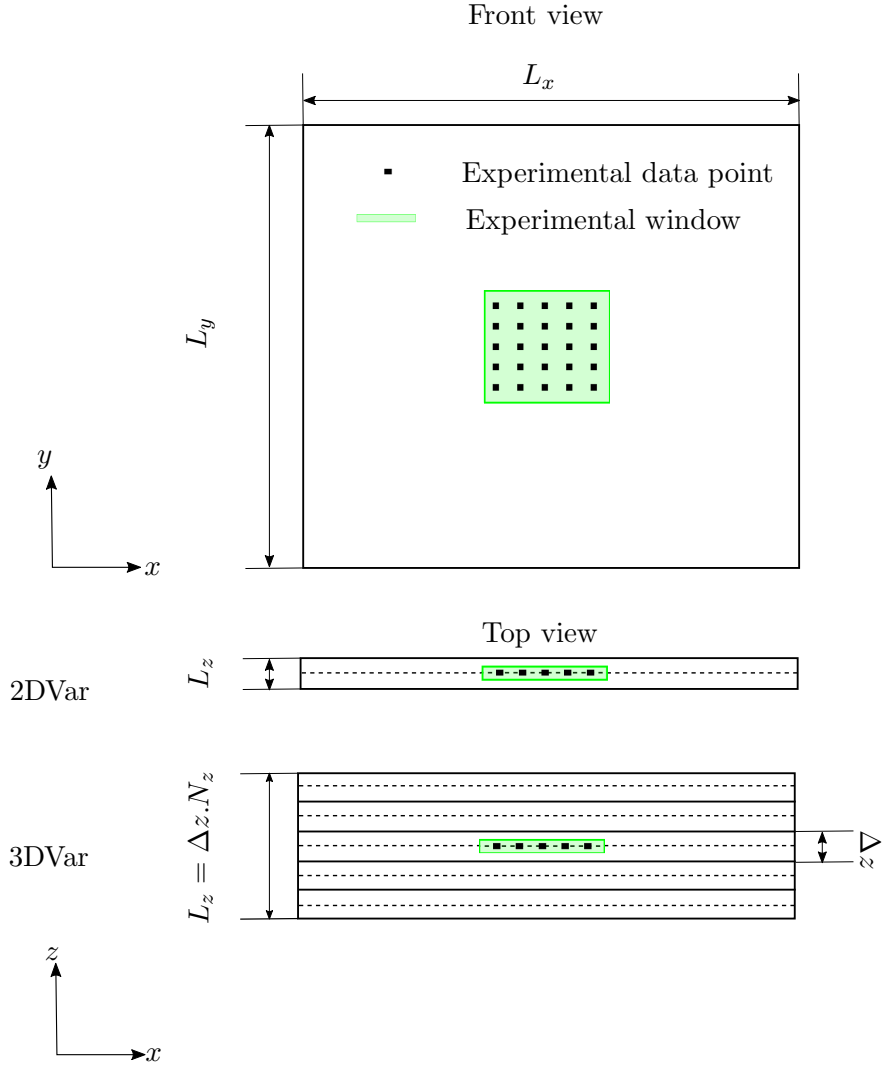}{%
\psfrag{h}[l]{Experimental data point}
\psfrag{i}[l]{Experimental window}
\psfrag{b}[c]{$L_x$}
\psfrag{a}[c]{$L_y$}
\psfrag{y}[c]{$y$}
\psfrag{z}[c]{$z$}
\psfrag{x}[c]{$x$}
\psfrag{e}[c]{$L_z$}
\psfrag{m}[c]{$L_z = \Delta z.N_z$}
\psfrag{c}[c]{2DVar}
\psfrag{k}[c]{$\Delta z$}
\psfrag{d}[c]{3DVar}
\psfrag{f}[c]{Top view}
\psfrag{g}[c]{Front view}
}
\caption{Schematic of the 2DVar and 3DVar setups, showing front and top views of the computational domain along with the experimental field of view and data points. Here, $L_x$, $L_y$, and $L_z$ are the streamwise, wall-normal, and spanwise domain lengths, respectively. The number of points in the spanwise direction is $N_z$, and the spanwise spacing of the computational mesh is $\Delta z$. Solid lines in the top view indicate a layer of cells, while dashed lines denote the position of the cell centroids.}
\label{fig:2D_3D_assm}
\end{figure}

Figure \ref{fig:2D_3D_assm} illustrates a generic setup for 2DVar and 3DVar. The computational domain is assumed to be significantly larger than the experimental field of view, with $L_x$ and $L_y$ defining its extent. The computational grid may be block-structured, structured, or unstructured. The experimental data points are arranged uniformly in the two spatial directions, consistent with measurements from techniques such as PIV. In 2DVar, the spanwise length $L_z$ is inconsequential as long as the number of points in the spanwise direction is $N_z=1$. This is evident from the top view in Fig. \ref{fig:2D_3D_assm}, where the cell centroids from the experimental data align with the computational grid, as indicated by the dashed lines.

The 3DVar setup differs in that the spanwise domain now extends by an integral multiple of the spanwise spacing $\Delta z$, given by $L_z = \Delta z N_z$, where $N_z > 1$. This effectively increases the number of divisions along the span for the same thickness. In Fig. \ref{fig:2D_3D_assm}, this is illustrated in the top view under the 3DVar section, where only one layer of cell centroids aligns with the experimental data points. The spanwise spacing $\Delta z$ is chosen to be slightly larger than the laser sheet thickness in a PIV setup and is defined by the limits $z_{max}$ and $z_{min}$ such that $\Delta z = z_{max} - z_{min}$, which is necessary for projection and smoothing operations described in Appendix. \ref{appendix:projection_smoothing}. In the schematic shown (Fig. \ref{fig:2D_3D_assm}), $N_z = 5$; however, for practical cases, the choice of $N_z$ depends on the specific problem. The spanwise domain must be sufficiently resolved to allow the development of the out-of-plane velocity component.

\section{Flow configuration and reference data}
\label{section:flow_configuration_and_reference_data}
In this section, details about flow over an airfoil in deep stall are provided. Although the experimental data are known in a single plane, the flow is highly three-dimensional, which motivates a comparison between 2DVar and 3DVar. Section \ref{subsection:flow_configuration} describes an airfoil in deep stall. The 2D validation reference data are described in Sec. \ref{subsection:2D_synthetic_data} and the experimental data are described in Sec.\ref{subsection:experimental_refernece_data}. The distinction between the two data sets is brought out by comparing their respective divergence errors. 

\subsection{Flow configuration}
\label{subsection:flow_configuration}
While we are interested in highly separated flows in general, we aim to demonstrate our methodology on a specific case of airfoil stall. The reason for this choice is the complex physics and flow features that are involved in the process of airfoil stall, including, but not limited to, separation (leading or trailing edge), presence of leading and trailing edge shear layers, and recirculating flow. Additionally, such strong streamline curvature and the presence of a pressure gradient make it challenging for existing turbulence models to capture the flow physics. We use a NACA0012 airfoil at an angle of attack $\alpha = 15^\circ$ and a chord-based Reynolds number $Re_c \approx 7.5 \times 10^4$, which is in the deep stall regime. 
 
\subsection{Two-dimensional synthetic data}\label{subsection:2D_synthetic_data}
We generate a validation dataset to demonstrate the assimilation of a flow field that satisfies 2D continuity. The reference data are generated using a $k -\omega$ SST \cite{menter1994two} (also referred to as the Menter SST) RANS turbulence model. This is a two-equation turbulence model that solves the transport equations for turbulent kinetic energy $k$ and specific dissipation rate $\omega$. The benefit of using this model is that it blends the $k - \epsilon$ \cite{jones1972prediction} model with the standard $k-\omega$ \cite{wilcox1988reassessment} (also known as the Wilcox $k -\omega$) model to overcome the limitations of using the two models in isolation. The RANS equations are solved using the Semi-Implicit Method for Pressure Linked Equations (SIMPLE) \cite{patankar1972calculation} through OpenFOAM's inbuilt \texttt{simpleFOAM} solver. The gradients are calculated using a second-order accurate central differencing scheme. All flow variables (velocity, $k$ and $\omega$) are discretized using a first-order upwind method to promote the stability of the solver. Each matrix equation is solved using the Gauss-Seidel method. Convergence is determined based on a tolerance of $10^{-5}$ for the residual of pressure and velocity components. 

\begin{figure}[h!]
     \centering
         \psfragfig[width=0.7\textwidth]{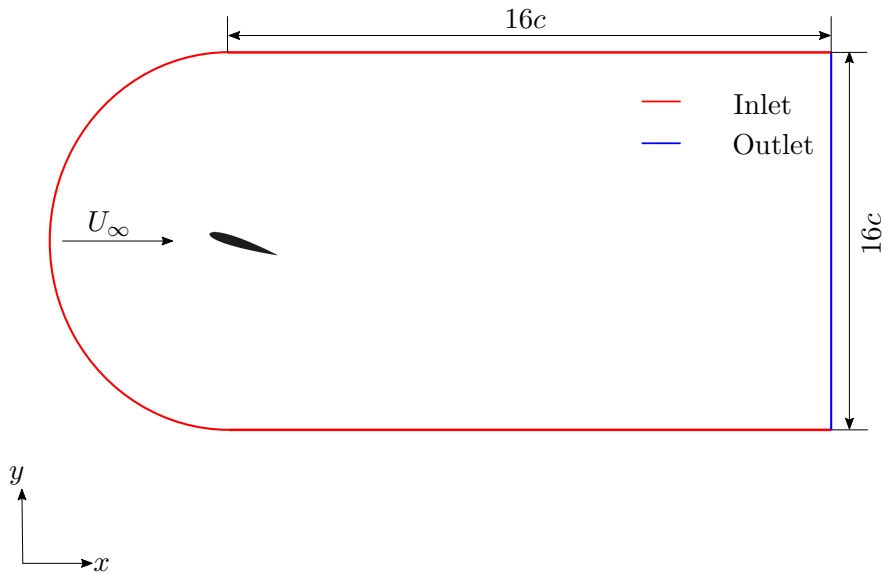}{%
    \psfrag{a}[c]{$U_{\infty}$}
    \psfrag{b}[c]{$16c$}
    \psfrag{c}[c]{$16c$}
    \psfrag{d}[c]{$0.1c$}
    \psfrag{f}[l]{Inlet}
    \psfrag{g}[l]{Outlet}
    \psfrag{h}[l]{Sides}
    \psfrag{x}[c]{$x$}
    \psfrag{y}[c]{$y$}
    \psfrag{i}[c]{Front view}
    \psfrag{j}[c]{Top view}

    }
    \caption{The schematic of the shape of the computational domain along with the inlet and outlet boundaries. } 
    \label{fig:domain}
\end{figure} 

The grid used for the study is a C-type grid shown in Fig. \ref{fig:domain}. The domain extends by $16c$ downstream and $8c$ upstream from the leading edge of the airfoil and $0.1c$ in the out-of-plane direction. Unless otherwise specified, the positive $x$ is designated as streamwise, the positive $y$ as wall-normal, and the positive $z$ as spanwise directions and is part of the right-handed coordinate system. The computational mesh consists of mostly hexahedral cells and has a refinement in the region close to the airfoil to ensure that the shear layers and wake are well resolved. Inflation layers are added to resolve the viscous sub-layer so that wall functions are not needed. The 2D grid is extruded by a single layer of cells in the spanwise direction such that $N_z = 1$ where $N_z$ is the number of points in the spanwise direction. This is required since all problems are inherently treated as 3D in OpenFOAM. We will solve the problem as 2D by enforcing the relevant boundary conditions. The discretization results in $400$ points along the airfoil and a total of $44687$ cells in the entire domain.   

\begin{table}
 \caption{\label{tab:bc_2D} Boundary conditions for 2D reference data}
 \begin{ruledtabular}
 \begin{tabular}{lccccc}
 & Inlet & Outlet & Front \& back & Airfoil \\
 \hline
 $U$  & 1 & $\frac{\partial U}{\partial n} = 0$ & $\hat{n.}\vec{U}=0$ \& $\frac{\partial U_\parallel}{\partial n} = 0$ & no slip \\
 $p$  & $\frac{\partial p}{\partial n} = 0$ & $0$ & $\frac{\partial p}{\partial n} = 0$ & $\frac{\partial p}{\partial n} = 0$ \\
 $k$  & $k_{farfield}$ & $\frac{\partial k}{\partial n} = 0$ & $\frac{\partial k}{\partial n} = 0$ & $0$ \\
$\omega$  & $\omega_{farfield}$ & $\frac{\partial \omega}{\partial n} = 0$ & $\frac{\partial \omega}{\partial n} = 0$ & $\omega_{wall}$ \\
 \end{tabular}
 \end{ruledtabular}
\end{table}

\begin{figure}[h!]
     \centering
         \psfragfig[width=0.7\textwidth]{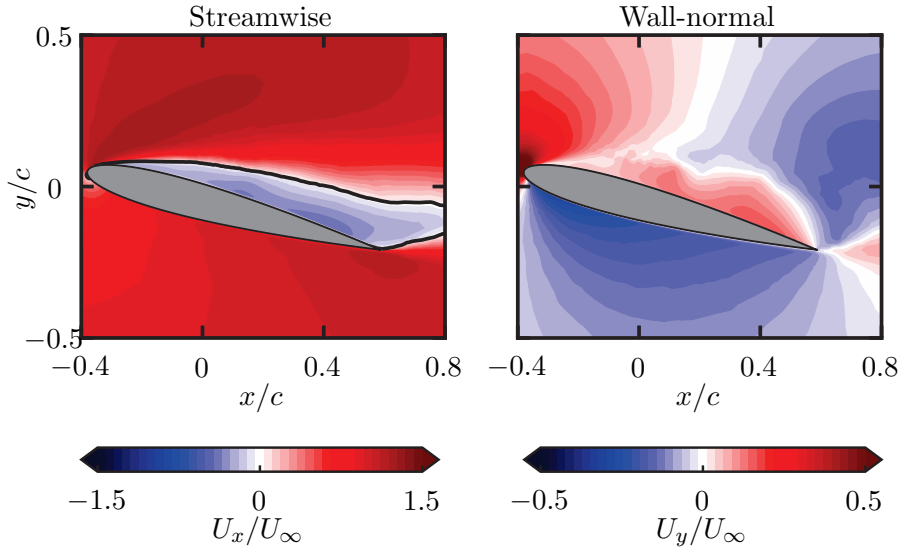}{%
    \psfrag{A}[c]{Streamwise}
    \psfrag{B}[c]{Wall-normal}
    \psfrag{D}[c]{$y/c$}
    \psfrag{E}[c]{$x/c$}
    \psfrag{f}[c]{$-1.5$}
    \psfrag{g}[c]{$-0.4$}
    \psfrag{h}[c]{$-0.5$}
    \psfrag{i}[c]{$0$}
    \psfrag{j}[c]{$0.5$}
    \psfrag{k}[c]{$0.4$}
    \psfrag{l}[c]{$1.5$}
    \psfrag{m}[c]{$U_x/U_{\infty}$}
    \psfrag{n}[c]{$U_y/U_{\infty}$}
    \psfrag{o}[c]{$0.8$}
    }
    \caption{Mean streamwise and wall-normal velocity fields obtained from $k-\omega$ SST RANS turbulence model (synthetic reference). The $U_x = 0$ contour is shown as solid black line. } 
    \label{fig:kOmega2D}
\end{figure}

The boundary conditions used to generate the dataset are summarized in Tab. \ref{tab:bc_2D}.  All quantities are normalized by the free stream velocity $U_\infty$ and the chord $c$. We employ a standard velocity inlet, pressure outlet boundary condition for incompressible flows where the velocity and pressure are set to fixed values at the inlet and outlet of the domain, respectively. The inlet boundary values for the turbulence variables $k$ and $\omega$ are set based on the freestream value recommendations from Ref. \cite{menter1994two} given by
\begin{align}
    \frac{U_{\infty}}{L} < \omega_{reference} < 10 \frac{U_{\infty}}{L}, \\
     \frac{10^{-5}U_{\infty}^2}{Re_L} < k_{reference} <   \frac{0.1U_{\infty}^2}{Re_L},
\end{align}
where $L$ is the approximate length of the domain and $Re_L$ is the Reynolds number based on this length. The front and back lateral faces of the domain are designated as symmetry boundaries in OpenFOAM for all quantities. This sets the outflow to be zero for normal components of vector quantities (such as velocity) and a zero gradient for tangential components, treating the problem as 2D. The airfoil surface is designated no slip for velocity and zero gradient for pressure. A fixed value is set for $k$ and the wall condition for $\omega$ is set to $\omega_{wall} = 60\nu/\beta_1(\Delta d_1)^2$ as per Ref. \cite{menter1994two} where $\Delta d_1$ is the distance to the first cell centroid from the wall and $\beta_1$ is an empirical constant. 

The mean streamwise and wall-normal velocity fields are presented in Fig.~\ref{fig:kOmega2D}. The domain for presenting the fields is selected to align with the experimental field of view, as described in Sec. \ref{subsection:experimental_refernece_data}, ensuring a fair comparison. The contour for \( U_x = 0 \) is also plotted to highlight the extent of the recirculation region. The flow field exhibits the characteristic behavior of an airfoil in deep stall, featuring leading-edge flow separation and a large recirculation zone. The use of a no-outflow boundary condition on the lateral faces of the domain ensures that the flow remains effectively two-dimensional, with negligible out-of-plane velocity components. A further verification of the two-dimensionality of the flow can be achieved by examining the magnitudes of the velocity gradients in the continuity equation, specifically \( \partial U_x / \partial x \), \( \partial U_y / \partial y \), and \( \partial U_z / \partial z \). The out-of-plane velocity gradient term \( \partial U_z / \partial z \) is on the order of \( 10^{-16} \), which is at the level of machine precision, indicating that any variation in the spanwise direction is negligible. Additionally, the overall divergence of the velocity field, computed from the in-plane terms as $1/V \int_V \nabla \cdot U \, dV \approx 1.73 \times 10^{-7}$, further confirms that the flow remains effectively two-dimensional.  

\subsection{Experimental reference data}
\label{subsection:experimental_refernece_data}
 The reference mean velocity fields are obtained from the experiment performed in Ref. \cite{carter2023low}. Time-resolved, planar PIV measurements are taken in a water flume facility. The test section is $6.75$ m long, $1.2$ m wide and $0.5$ m deep. The airfoil has a chord of $15$ cm and a span of $70$ cm. More than half the span is submerged. Three $4$-megapixel, high-speed Phantom Veo $640$-S cameras are used to capture the images. The forces are measured using a six-axis force/torque load cell (ATI Delta IP65) mounted to the airfoil and synchronized to the PIV acquisition using an NI USB-$6251$ data acquisition device (DAQ). The acquisition frequency is $1$kHz. Multipass PIV is performed using a verified in-house MATLAB code with $3$ passes per window size and square windows decreasing from 64 by 64 pixels to 32 by 32 pixels to 24 by 24 pixels with $50\%$ overlap. This results in $197 \times 249$ vectors with $181$ vectors along the airfoil chord. Further details can be found in Ref. \cite{carter2023low}. A generous mask up to $y^+ \approx 100$ is used, removing the data close to the airfoil surface that is corrupted from surface reflections.
 
\begin{figure}[h!]
     \centering
         \psfragfig[width=0.7\textwidth]{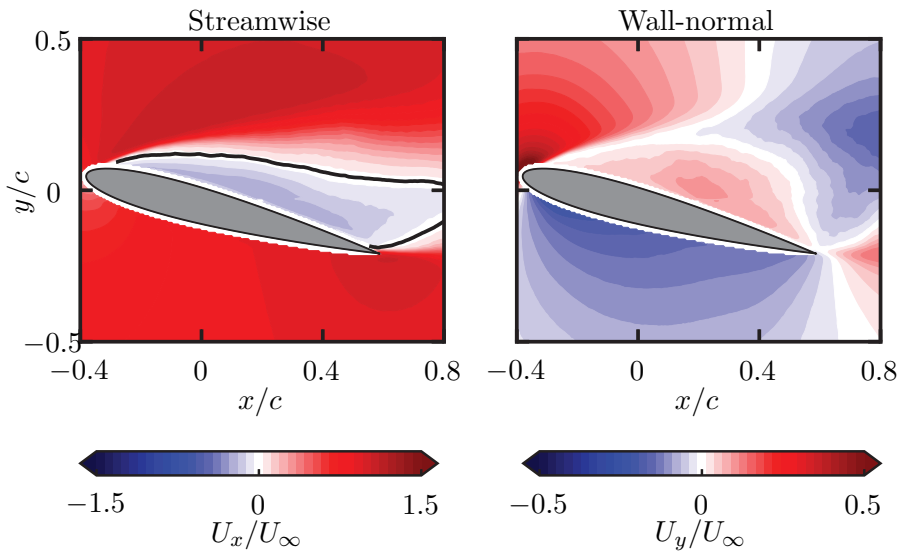}{%
    \psfrag{A}[c]{Streamwise}
    \psfrag{B}[c]{Wall-normal}
    \psfrag{D}[c]{$y/c$}
    \psfrag{E}[c]{$x/c$}
    \psfrag{f}[c]{$-1.5$}
    \psfrag{g}[c]{$-0.4$}
    \psfrag{h}[c]{$-0.5$}
    \psfrag{i}[c]{$0$}
    \psfrag{j}[c]{$0.5$}
    \psfrag{k}[c]{$0.4$}
    \psfrag{l}[c]{$1.5$}
    \psfrag{m}[c]{$U_x/U_{\infty}$}
    \psfrag{n}[c]{$U_y/U_{\infty}$}
    \psfrag{o}[c]{$0.8$}
    }
    \caption{Mean streamwise and wall-normal velocity fields obtained from experiment. The $U_x = 0$ contour is shown as solid black line.} 
    \label{fig:doug2D}
\end{figure}

The experimental mean streamwise and wall-normal velocity fields are shown in Fig. \ref{fig:doug2D}. Just as in the synthetic reference data case, the $U_x = 0$ contour is plotted to delineate the region of reverse flow. The mean streamwise velocity field is representative of a deep stall case with leading edge separation and a large recirculation bubble. This is a case of well-resolved flow field but the field of view (FOV) is limited. The streamwise extent of the FOV is about $1.2c$ while the wall-normal extent is $1c$. The full extent of the recirculation bubble is not captured. 

\begin{figure}[h!]
     \centering
         \psfragfig[width=0.4\textwidth]{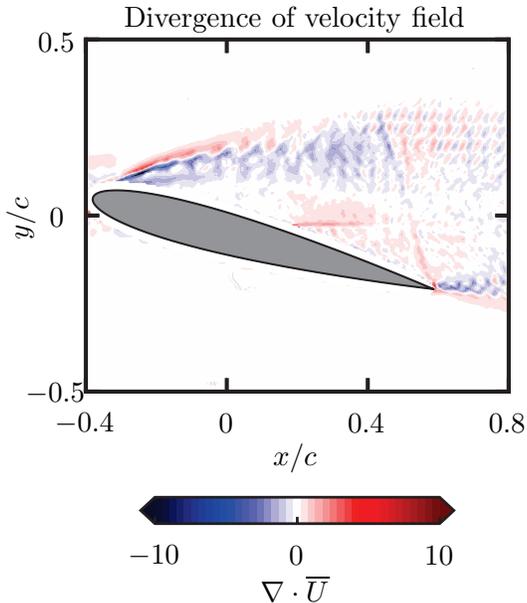}{%
    \psfrag{A}[c]{Divergence of velocity field}
    \psfrag{D}[c]{$y/c$}
    \psfrag{E}[c]{$x/c$}
    \psfrag{f}[c]{$-10$}
    \psfrag{g}[c]{$-0.4$}
    \psfrag{h}[c]{$-0.5$}
    \psfrag{i}[c]{$0$}
    \psfrag{j}[c]{$0.5$}
    \psfrag{k}[c]{$0.4$}
    \psfrag{l}[c]{$10$}
    \psfrag{m}[c]{$\nabla \cdot\overline{U}$}
    \psfrag{o}[c]{$0.8$}
    }
    \caption{The divergence of the experimental reference data.} 
    \label{fig:doug_divergence}
\end{figure} 

A key difference between the experimental dataset and the synthetic one is the presence of divergence errors. We only have access to the streamwise and wall-normal components of velocity in this experimental dataset. As a result, the divergence of the velocity field would lack the contribution of the out-of-plane velocity gradient (in this case, $\partial U_z/\partial z$). The 2D divergence field of this experimental dataset can then be calculated using the streamwise and wall-normal velocity gradients only as
\begin{align}
    \label{eq:2D_divergence}
    \nabla \cdot U = \partial U_x/\partial x + \partial U_y/\partial y.
\end{align}
Since the data are present on a uniform grid, a central differencing scheme (accurate up to second order) is used to compute the 2D divergence, which is presented in Fig. \ref{fig:doug_divergence}. There are large values in the leading edge shear layer. Additionally, there is an unphysical artifact from the trailing edge of the airfoil that is caused due to the overlap of the laser sheets from the pressure and suction sides of the airfoil. The divergence in the wake of the airfoil (that coincides with the region of high experimental uncertainty) also appears grainy. All these features are quite typical of an experimental dataset. The 2D divergence of the dataset on a discrete grid is then calculated as,
\begin{align}
\label{eq:divergence}
\nabla \cdot U = \frac{1}{N_e} \sum_{N_e} \left(\frac{\partial U_x}{\partial x} + \frac{\partial U_y}{\partial y}\right).   
\end{align}
The magnitude of the 2D divergence is $\nabla.U = -0.0359$ which is five orders of magnitude larger than the divergence of the synthetic data field in Sec. \ref{subsection:2D_synthetic_data}. The use of this dataset for 2DVar assumes that the velocity field obeys 2D continuity. It is evident from Fig. \ref{fig:doug_divergence} and the magnitude of 2D divergence reported in Tab. \ref{tab:divergence} that this is not the case. In reality, the presence of noise would make it challenging to achieve a truly divergence-free velocity field. However, such a large discrepancy could only be explained by the three-dimensionality of the flow field. This would require the contribution of the out-of-plane velocity gradient to have a fully divergence-free velocity field within experimental uncertainty.  

\begin{table}
    \caption{\label{tab:divergence} Divergence values of synthetic and experimental reference data.}
    \begin{ruledtabular}
    \renewcommand{\arraystretch}{1.1} 
    \begin{tabular}{@{}l c c@{}} 
        & Synthetic & Experimental  \\ 
        \hline
        Divergence  & $-1.73 \times 10^{-7}$ & $-0.0359$ \\  
    \end{tabular}
    \end{ruledtabular}
\end{table}

\section{Baseline computation}
\label{section:baseline_computation} 
With the reference datasets defined and characterized in Sec. \ref{section:flow_configuration_and_reference_data}, we present the details of baseline simulations. The choice of the turbulence model, computational mesh, and boundary conditions for 2DVar and 3DVar assimilation are discussed in Sec. \ref{subsection:setup_baseline_computation}. The results from the baseline computation are then compared with the reference data in Sec. \ref{subsection:results_baseline_computation}.

\subsection{Setup for baseline computation}
\label{subsection:setup_baseline_computation}

We use the Spalart-Allmaras \cite{spalart1992one} (SA) turbulence model to close the set of equations for the baseline model. It applies the Boussinesq approximation through the eddy viscosity, which is obtained by solving the transport equation for a surrogate variable $\Tilde{\nu}$. The functional form of the turbulence transport equation is given by:
\begin{align}
\label{equation:samodel}
    \frac{D \Tilde{\nu}}{D t} = P(\Tilde{\nu}, \mathbf{w}) + T(\Tilde{\nu}, \mathbf{w}) - D(\Tilde{\nu}, \mathbf{w}), 
\end{align}
where $\mathbf{w}$ is the vector of state variables such as mean velocity, pressure, and momentum flux, and $P$, $T$, and $D$ are the production, transport, and cross-diffusion terms, respectively. The details of the mesh, the solver, choice of parameters such as kinematic viscosity $\nu$, chord length $c$ and freestream velocity $U_{\infty}$ can be found in Sec. \ref{subsection:2D_synthetic_data}. We study three separate cases: 2DVar assimilation of synthetic and experimental reference data, and 3DVar assimilation of experimental data. Regardless of the case, all  baseline computations use the SA model by setting the corrective forcing term $f_{c_i} = \mathbf{0}$ in Eq. \ref{equation:mom_correction}. 

\begin{table}
 \caption{\label{tab:bc_bsl} Boundary conditions for baseline SA model}
 \begin{ruledtabular}
 \begin{tabular}{lccccc}
 & Inlet & Outlet & Front \& back & Airfoil \\
 \hline
 $\tilde{\nu}$  & $\tilde{\nu}_{\infty}$ & $\frac{\partial \tilde{\nu}}{\partial n} = 0$ &  $\frac{\partial \tilde{\nu}}{\partial n} = 0$ & $\tilde{\nu} = 0$ \\
  $\nu_t$  & \texttt{calculated} & \texttt{calculated} &  $\frac{\partial \nu_t}{\partial n} = 0$ & \texttt{calculated} \\
 \end{tabular}
 \end{ruledtabular}
\end{table}

The baseline computations for 2DVar are performed on the same computational mesh as defined in Sec. \ref{subsection:2D_synthetic_data} using the C-type grid as shown in Fig. \ref{fig:domain}. The boundary conditions for velocity and pressure are the same as those used to generate the synthetic reference data, summarized in Tab. \ref{tab:bc_2D}. The boundary conditions for the eddy viscosity $\nu_t$ and modified eddy viscosity $\tilde{\nu}$ are summarized in Tab. \ref{tab:bc_bsl}. We set a fixed value inlet boundary for $\tilde{\nu}_{\infty} = 3\nu_{\infty}$ to $5\nu_{\infty}$ with the recommendation from Ref. \cite{spalart1992one} and a fixed gradient outlet boundary. The lateral faces of the domain are designated as a no-flow condition. The presence of inflation layers and the resolution of the viscous sub-layer means that there are no wall functions required, setting a value of $\tilde{\nu} = 0$ at the wall. Since the transport equations are solved for $\tilde{\nu}$ as outlined in Eq. \ref{equation:samodel}, the boundary condition for the eddy viscosity used to calculate the closure term is designated as \texttt{calculated} in OpenFOAM that estimates the value of $\nu_t$ based on a relation provided in Ref. \cite{spalart1992one}. 

\begin{figure}[h!]
     \centering
         \psfragfig[width=0.8\textwidth]{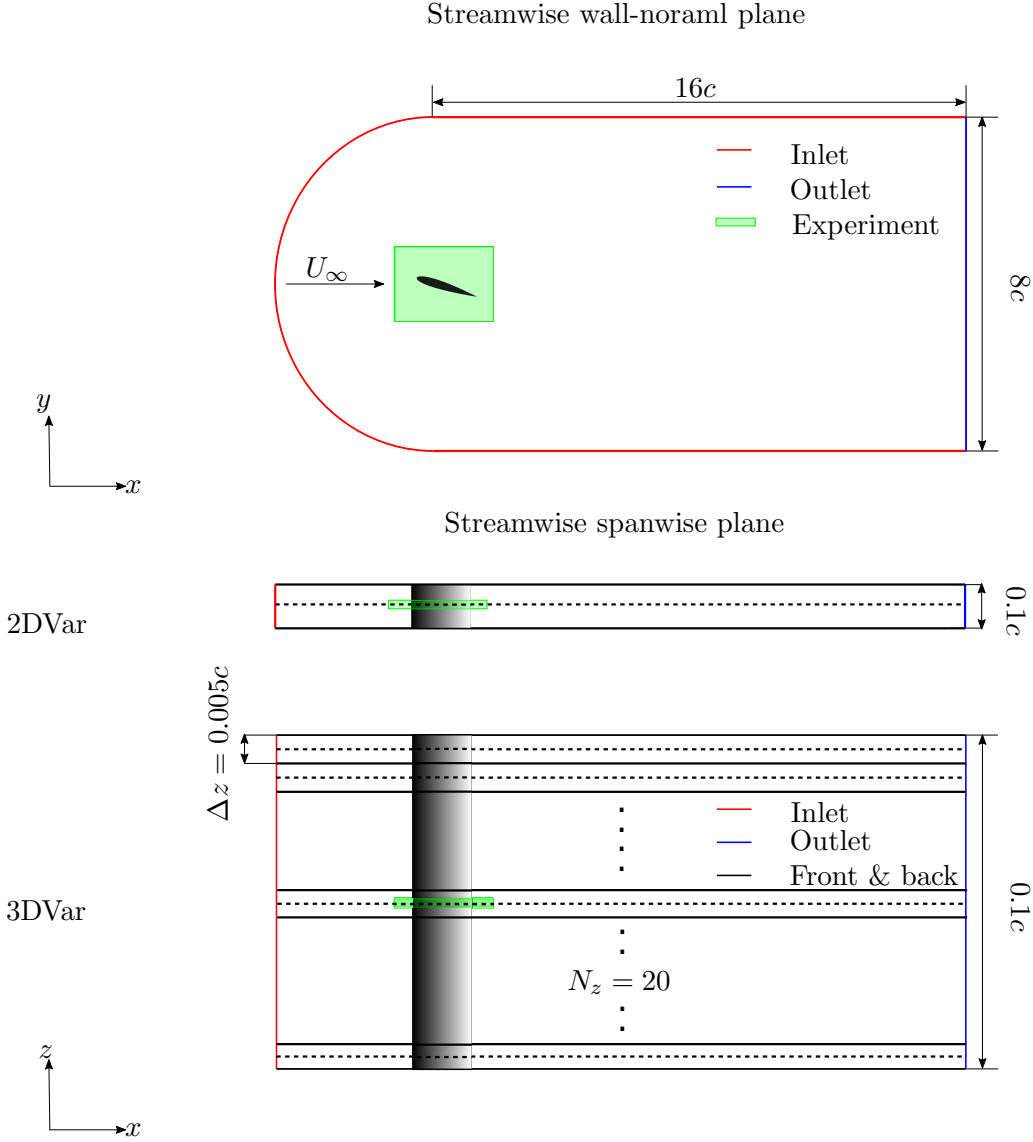}{%
    \psfrag{a}[c]{$U_{\infty}$}
    \psfrag{b}[c]{$16c$}
    \psfrag{c}[c]{2DVar}
    \psfrag{d}[c]{3DVar}
    \psfrag{q}[c]{$8c$}
    \psfrag{g}[l]{Outlet}
    \psfrag{f}[l]{Inlet}
    \psfrag{h}[l]{Front \& back}
    \psfrag{p}[c]{$0.1c$}
    \psfrag{i}[c]{Streamwise wall-noraml plane}
    \psfrag{j}[c]{Streamwise spanwise plane}
    \psfrag{y}[c]{$y$}
    \psfrag{x}[c]{$x$}
    \psfrag{z}[c]{$z$}
    \psfrag{k}[l]{Experiment}
    \psfrag{w}[c]{$\Delta z = 0.005c$}
    \psfrag{s}[c]{$N_z = 20$}
    }
    \caption{The computational domains for 2DVar and 3DVar are shown, highlighting the streamwise-wall-normal and streamwise-spanwise planes. For clarity, the spanwise length $L_z = 0.1c$ has been exaggerated in the 3DVar representation to better illustrate the divisions. The inlet, outlet, and lateral faces of the domain are labeled, along with the flow direction. The experimental field of view is also indicated, and the cell centroids are represented by dashed lines.} 
    \label{fig:2D_3D_var}
\end{figure}

The mesh used for 3DVar is different from that used for 2DVar. As described in Sec. \ref{subsection:data_assimilation_setup}, the computational grid for 3DVar can be generated by increasing the number of divisions along the span to $N_z = 20$ with the thickness of each layer in the spanwise direction being $\Delta z = 0.005c$. This results in a computational mesh size of $893,740$ cells. A schematic of this setup, showing the spanwise resolution, can be seen in Fig. \ref{fig:2D_3D_var}. The boundary conditions are the same as those summarized in Tabs. \ref{tab:bc_2D} and \ref{tab:bc_bsl}. The only difference is that the spanwise direction has sufficient resolution to allow the development of an out-of-plane velocity component that allows the enforcement of 3D continuity.  

The discrepancy between the baseline and reference mean velocity data will be quantified by using the $L_1$ norm field which is computed as
\begin{align}
\label{eqn:L1NormPaper}
    L_1 = \frac{|\mathcal{Q}(U_x) - \hat{U}_x| + |\mathcal{Q}(U_y) - \hat{U}_y|}{U_{\infty}},
\end{align}
where $\hat{U}_x$, $\hat{U}_y$ are the reference data mean streamwise and wall-normal velocities, respectively, and $\mathcal{Q}$ is the projection operator which depends on the reference data set. This projection operator allows the computation of the $L_1$ norm on the experimental grid. 

\subsection{Results from baseline computation}
\label{subsection:results_baseline_computation}

\begin{figure}[h!]
     \centering
         \psfragfig[width=1\textwidth]{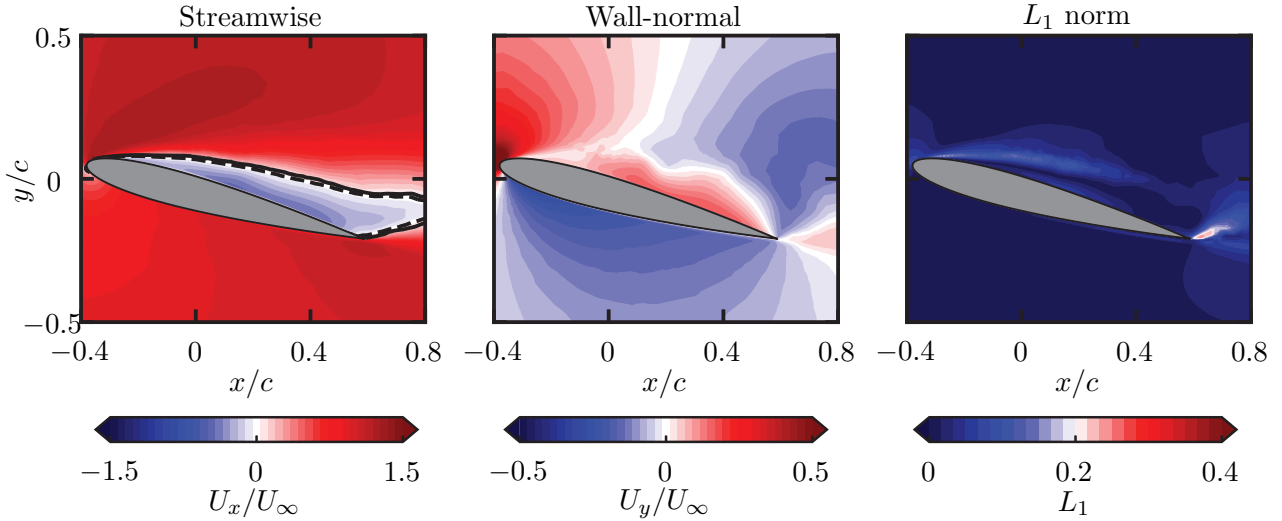}{%
    \psfrag{A}[c]{Streamwise}
    \psfrag{B}[c]{Wall-normal}
    \psfrag{D}[c]{$y/c$}
    \psfrag{E}[c]{$x/c$}
    \psfrag{f}[c]{$-1.5$}
    \psfrag{g}[c]{$-0.4$}
    \psfrag{h}[c]{$-0.5$}
    \psfrag{i}[c]{$0$}
    \psfrag{j}[c]{$0.5$}
    \psfrag{k}[c]{$0.4$}
    \psfrag{y}[c]{$0.4$}
    \psfrag{o}[c]{$0.8$}
    \psfrag{l}[c]{$1.5$}
    \psfrag{m}[c]{$U_x/U_{\infty}$}
    \psfrag{n}[c]{$U_y/U_{\infty}$}
    \psfrag{p}[c]{$L_1$}
    \psfrag{z}[c]{$1$}
    \psfrag{x}[c]{$0.2$}
        \psfrag{C}[c]{$L_1$ norm}
    }
    \caption{From left to right: mean streamwise velocity, mean wall-normal velocity fields, and the $L_1$ norm comparing mean velocity fields from the baseline SA RANS turbulence model with the reference synthetic data. The $U_x = 0$ contour is depicted as a solid black line for the synthetic reference data and as a dashed black line for the baseline model in the mean streamwise velocity field.} 
    \label{fig:bsl_komega_comp}
\end{figure}

We present the mean streamwise and wall-normal velocity fields computed using the baseline SA model in Fig. \ref{fig:bsl_komega_comp}. We offer a comparison of these fields with the synthetic reference dataset by computing the $L_1$ norm that is defined in Eq. \ref{eqn:L1NormPaper}. We use $\mathcal{Q} = 1$ since the synthetic reference data and baseline model computation share the same computational grid. The extent of recirculation is demarcated by the $U_x = 0$ contour which is a solid line for the reference data  and a dashed line for the baseline model. It is evident that the recirculation bubble sizes are not very different; this is to be expected since both computations employ RANS turbulence models that use the Boussinesq approximation. The differentiating point is that the baseline model is a one-equation turbulence model while the reference data are generated using a two-equation turbulence model. This difference is brought out as a discrepancy in the leading and trailing edge shear layers as seen in the plot of the $L_1$ norm field in Fig. \ref{fig:bsl_komega_comp}. The discrepancy that is noticed will aim to be reduced by performing 2DVar using this baseline model for the reference dataset. 

\begin{figure}[h!]
     \centering
         \psfragfig[width=1\textwidth]{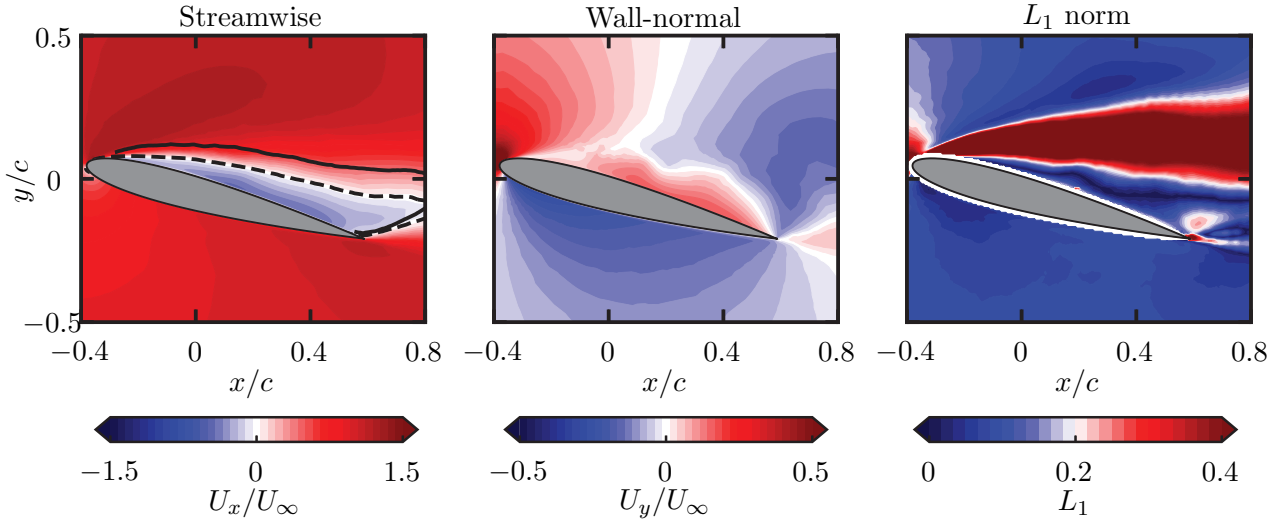}{%
    \psfrag{A}[c]{Streamwise}
    \psfrag{B}[c]{Wall-normal}
    \psfrag{D}[c]{$y/c$}
    \psfrag{E}[c]{$x/c$}
    \psfrag{f}[c]{$-1.5$}
    \psfrag{g}[c]{$-0.4$}
    \psfrag{h}[c]{$-0.5$}
    \psfrag{i}[c]{$0$}
    \psfrag{j}[c]{$0.5$}
    \psfrag{k}[c]{$0.4$}
    \psfrag{o}[c]{$0.8$}
    \psfrag{x}[c]{$0.2$}
    \psfrag{l}[c]{$1.5$}
    \psfrag{m}[c]{$U_x/U_{\infty}$}
    \psfrag{n}[c]{$U_y/U_{\infty}$}
    \psfrag{p}[c]{$L_1$}
    \psfrag{z}[c]{$1$}
        \psfrag{C}[c]{$L_1$ norm}
    }
    \caption{From left to right: mean streamwise velocity, mean wall-normal velocity fields, and the $L_1$ norm comparing mean velocity fields from the baseline SA RANS turbulence model with the reference experimental data. The $U_x = 0$ contour is depicted as a solid black line for the synthetic reference data and as a dashed black line for the baseline model in the mean streamwise velocity field.} 
    \label{fig:doug2D_bsl_comp}
\end{figure}

The comparison of the mean streamwise and wall-normal velocity field obtained from the baseline model with experimental reference data is presented in Fig. \ref{fig:doug2D_bsl_comp}. As with the previous case, we compare the extent of recirculation using the $U_x = 0$ contour where the baseline model is represented as a dashed line and the experimental data as a solid one, and also plot the $L_1$ norm. The experimental data are on a structured grid with a uniform distribution of points, whereas the baseline model is applied on a computational mesh that is unstructured. We use Eq. \ref{eqn:L1NormPaper} to compute the $L_1$ norm where the operator $\mathcal{Q}$ is linear interpolation to transfer the mean velocity data from the computational grid to the experimental one. The $L_1$ norm field reveals a very large discrepancy in the leading edge shear layer also corroborated by a reduced recirculation bubble observed in the baseline model. This is quite a large discrepancy compared to the one observed for synthetic reference data in Fig. \ref{fig:bsl_komega_comp} and is a clear indication that RANS turbulence models lack the accuracy required to model high Reynolds number airfoils in stall. 

The baseline model used for 2DVar solves the governing equations in 2D. The baseline model results for 3DVar are not shown here, as the streamwise and wall-normal velocity components obtained from applying the baseline model on a 3D computational mesh do not differ significantly from those in 2DVar. The spanwise component of the mean velocity, computed using the baseline model, is negligibly small (on the order of $10^{-7}$). While the computational mesh is sufficiently resolved along the span, limitations of the model itself may prevent accurate resolution of the spanwise velocity component.

\section{Assimilation of synthetic data}
\label{section:assimilation_of_synthetic_data}

Within this section, we examine the assimilation of the mean velocity field using synthetic reference data. Our analysis centers on the momentum forcing as the control variable and the insights derived from it. We particularly focus on the consequence of applying 2D continuity and how that impacts the magnitude of the forcing term. 

We perform 2DVar assimilation of synthetic data as described in Sec. \ref{subsection:2D_synthetic_data} using the SA model as baseline. We do not perform any projection or smoothing operations and, as such, $\mathcal{Q} = 1$ in Eq. \ref{equation:mom_source_obj_func}. The objective function drops by $\approx 80\%$ of its initial value, suggesting that the algorithm is performing its task as intended. The divergence of the assimilated mean velocity field is calculated to be $\nabla \cdot U \approx 2 \times 10^{-7}$ using the streamwise and wall-normal gradients only.  

\begin{figure}[h!]
     \centering
         \psfragfig[width=1\textwidth]{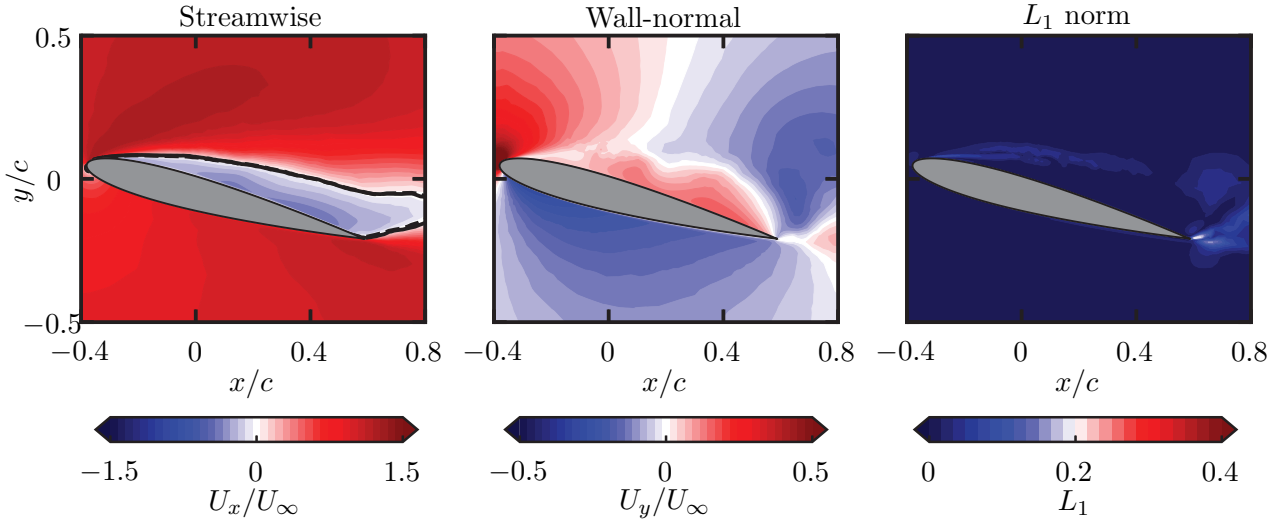}{%
    \psfrag{A}[c]{Streamwise}
    \psfrag{B}[c]{Wall-normal}
    \psfrag{D}[c]{$y/c$}
    \psfrag{E}[c]{$x/c$}
    \psfrag{f}[c]{$-1.5$}
    \psfrag{g}[c]{$-0.4$}
    \psfrag{h}[c]{$-0.5$}
    \psfrag{i}[c]{$0$}
    \psfrag{x}[c]{$0.2$}
    \psfrag{y}[c]{$0.4$}
    \psfrag{j}[c]{$0.5$}
    \psfrag{k}[c]{$0.4$}
    \psfrag{o}[c]{$0.8$}
    \psfrag{x}[c]{$0.2$}
    \psfrag{l}[c]{$1.5$}
    \psfrag{m}[c]{$U_x/U_{\infty}$}
    \psfrag{n}[c]{$U_y/U_{\infty}$}
    \psfrag{p}[c]{$L_1$}
    \psfrag{z}[c]{$1$}
        \psfrag{C}[c]{$L_1$ norm}
    }
    \caption{From left to right: 2DVar-assimilated mean streamwise velocity, mean wall-normal velocity fields, and the $L_1$ norm comparing the assimilated mean velocity fields with the reference synthetic data. The $U_x = 0$ contour is depicted as a solid black line for the synthetic reference data and as a dashed black line for the data-assimilated model.} 
    \label{fig:assm_komega_comp}
\end{figure}

The data-assimilated mean streamwise and wall-normal velocity fields are presented in Fig. \ref{fig:assm_komega_comp}. The discrepancy in the leading edge shear layer is smaller and this is evident from the $L_1$ norm field and the size of the recirculation region (comparing the reference and assimilated $U_x = 0$ contours). Although the improvement in the recirculation bubble size is only marginal, it shows a much better agreement with the reference data. 

\begin{figure}[h!]
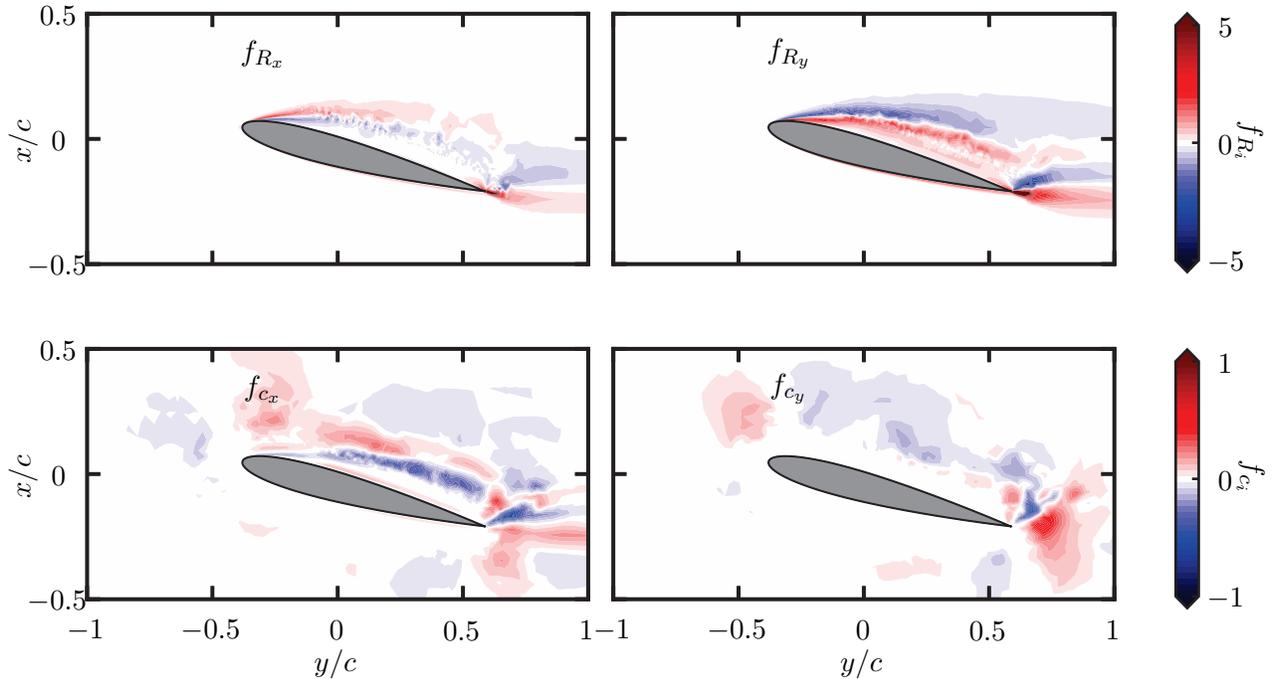

     \centering
         \psfragfig[width=1\textwidth]{komega_compare_forcing}{%
    \psfrag{A}[c]{$f_{R_x}$}
    \psfrag{B}[c]{$f_{R_y}$}
    \psfrag{C}[c]{$f_{c_x}$}
    \psfrag{W}[c]{$f_{c_y}$}
    \psfrag{E}[c]{$y/c$}
    \psfrag{D}[c]{$x/c$}
    \psfrag{g}[c]{$-0.4$}
    \psfrag{h}[c]{$-0.5$}
    \psfrag{i}[c]{$0$}
    \psfrag{j}[c]{$0.5$}
    \psfrag{k}[c]{$0.4$}
    \psfrag{o}[c]{$0.5$}
    \psfrag{n}[c]{$-0.4$}
    \psfrag{x}[c]{$-0.5$}
    \psfrag{p}[c]{$1$}
    \psfrag{n}[c]{$-1$}
    \psfrag{z}[c]{$5$}
    \psfrag{s}[c]{$-5$}
    \psfrag{y}[c]{$f_{R_i}$}
    \psfrag{w}[c]{$f_{c_i}$}
    }
    \caption{Comparison of the streamwise and wall-normal components of the corrective forcing $f_{c_i}$ with the Reynolds forcing $f_{R_i}$ for the 2DVar assimilation using synthetic reference data.} 
    \label{fig:komega_compare_forcing}
\end{figure}

While the agreement between the assimilated and reference mean velocity fields is encouraging, we need to examine other quantities of interest such as Reynolds stresses and the control variable to derive more insight into the 2DVar assimilation. As shown in Eq. \ref{equation:mom_correction}, we define two forcing terms, a Reynolds forcing term $f_{R_i} = \nabla \cdot R_{ij}$ where $R_{ij}$ is the Reynolds stress tensor, and the corrective forcing $f_{c_i}$ which is optimized to reduce the mean velocity discrepancy. We present a comparison of the corrective forcing with the Reynolds forcing for both streamwise and wall-normal components in Fig. \ref{fig:komega_compare_forcing}. The Reynolds forcing acts along the leading and trailing edge shear layers, which is expected for a case of leading edge flow separation. The streamwise and wall-normal components of the corrective forcing are also presented in the same figure. The majority of the corrective forcing acts in the regions that have high discrepancy, as shown in the $L_1$ norm field in Fig. \ref{fig:bsl_komega_comp}. Regions of positive forcing are complemented with negative forcing so as to ensure that the forcing field, as a whole, is divergence-free.

The maximum and minimum values of the components of corrective and Reynolds forcing are presented in Tab. \ref{tab:forcing}. A key observation is that the streamwise and wall-normal components of the Reynolds forcing are at least an order of magnitude greater than their corrective forcing counterparts. To explain this difference, we need to first investigate the action of the forcing term. Reynolds forcing arises from the eddy viscosity and the mean velocity gradient, effectively representing the contribution of the turbulence model. Any change in the mean velocity, induced by the corrective forcing, influences the turbulence variable (in this case, the eddy viscosity $\nu_t$), which in turn affects the magnitude of the Reynolds stresses. The forcing term does not distinguish between different sources of discrepancies between the reference data and the baseline model—it solely compensates for these discrepancies. We have deliberately chosen identical parameters for generating the synthetic reference data and applying the baseline model. A crucial consideration is that both velocity fields satisfy the 2D continuity equation, ensuring that the forcing term does not compensate for any deviation from a divergence-free field. The only distinction lies in the formulation of the turbulence model. Consequently, it is reasonable to expect that the corrective forcing primarily accounts for the differences between these models.

\begin{table}
    \caption{\label{tab:forcing} Maximum and minimum values for forcing components in 2DVar}
    \begin{ruledtabular}
    \renewcommand{\arraystretch}{1.3} 
    \begin{tabular}{p{3cm} S S S S} 
        & \multicolumn{2}{l}{\textbf{Corrective Forcing}} & \multicolumn{2}{l}{\textbf{Reynolds Forcing}} \\ 
        & {$f_{c_x}$} & {$f_{c_y}$} & {$f_{R_x}$} & {$f_{R_y}$} \\ 
        \hline
         Minimum & -0.83 & -0.77 & -8.74 & -6.74 \\ 
        Maximum & 0.38 & 0.65 & 7.89 & 10.33 \\   
    \end{tabular}
    \end{ruledtabular}
\end{table}

Since the corrective forcing accounts for the differences between the models, its lower magnitude compared to the Reynolds forcing suggests that the baseline model can capture most of the physics present in the reference data by simply modifying the eddy viscosity. This outcome is expected, as both the SA and $k-\omega$ SST turbulence models rely on the Boussinesq approximation, which relates the Reynolds stresses to the mean strain rate tensor through an eddy viscosity field. As mentioned earlier, this holds true when the inference of the corrective forcing field is unambiguous.

\begin{figure}[h!]
     \centering
         \psfragfig[width=0.85\textwidth]{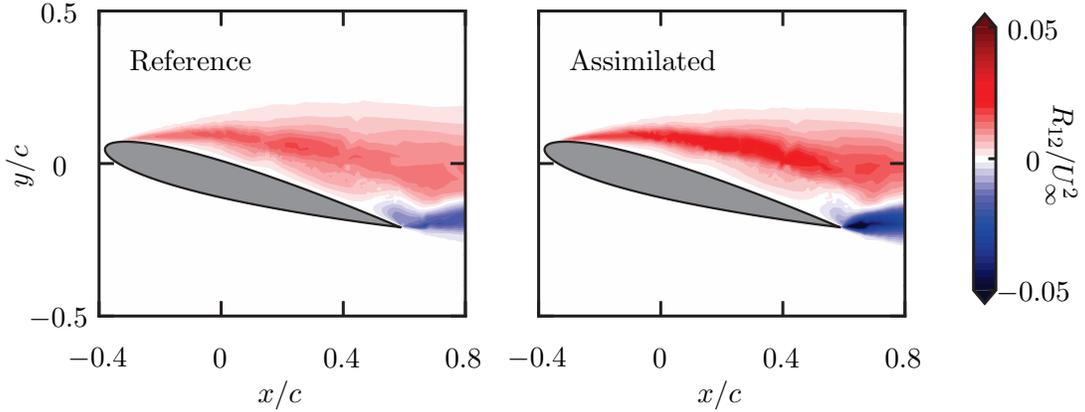}{%
    \psfrag{A}[l]{Reference}
    \psfrag{B}[l]{Assimilated}
    \psfrag{D}[c]{$y/c$}
    \psfrag{E}[c]{$x/c$}
    \psfrag{n}[c]{$-0.4$}
    \psfrag{h}[c]{$-0.5$}
    \psfrag{i}[c]{$0$}
    \psfrag{j}[c]{$0.5$}
    \psfrag{z}[c]{$0.05$}
    \psfrag{o}[c]{$0.4$}
    \psfrag{p}[c]{$0.8$}
    \psfrag{x}[c]{$-0.05$}
    \psfrag{m}[c]{$R_{12}/U_{\infty}^2$}
    }
    \caption{Reynolds shear stress from reference $k-\omega$ SST data and 2DVar assimilation.} 
    \label{fig:komega_rss}
\end{figure}
The small magnitudes of the corrective forcing, as seen in Tab. \ref{tab:forcing}, are reflected in the strong agreement of the Reynolds shear stresses between the reference and assimilated data, as shown in Fig. \ref{fig:komega_rss}. The shape of the Reynolds shear stress profiles is preserved, with only minor differences in magnitude observed between the reference $k-\omega$ SST and assimilated data.

\begin{table}[h!]
    \caption{\label{tab:komega_pressure} Pressure coefficient errors for 2DVar using synthetic reference}
    \begin{ruledtabular}
    \renewcommand{\arraystretch}{1.1} 
    \begin{tabular}{@{}l c   c@{}} 
        & $Error_b$ &  $Error_a$  \\ 
        \hline
        Magnitude & $0.008$ & $0.004$ \\  
    \end{tabular}
    \end{ruledtabular}
\end{table}

\begin{figure}[h!]
     \centering
         \psfragfig[width=0.8\textwidth]{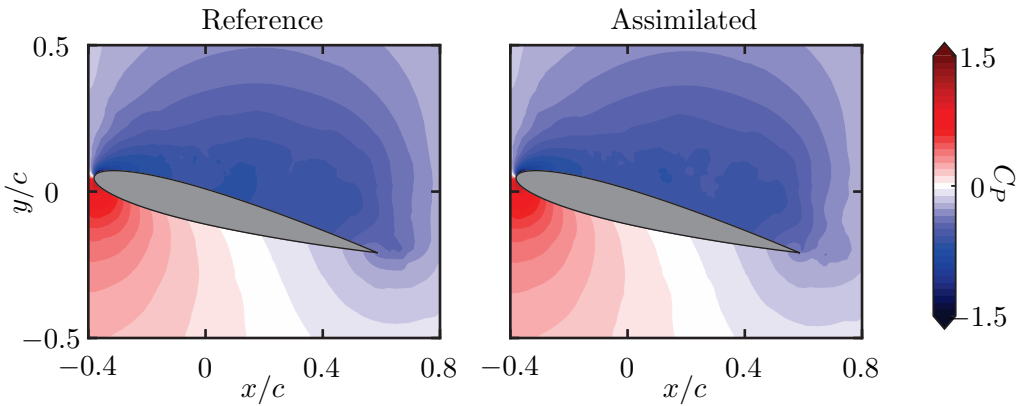}{%
    \psfrag{A}[c]{Reference}
    \psfrag{B}[c]{Assimilated}
    \psfrag{D}[c]{$y/c$}
    \psfrag{E}[c]{$x/c$}
    \psfrag{f}[c]{$-1.5$}
    \psfrag{g}[c]{$-0.4$}
    \psfrag{h}[c]{$-0.5$}
    \psfrag{i}[c]{$0$}
    \psfrag{r}[c]{$-1$}
    \psfrag{x}[c]{$0.1$}
    \psfrag{y}[c]{$0.2$}
    \psfrag{j}[c]{$0.5$}
    \psfrag{k}[c]{$0.4$}
    \psfrag{o}[c]{$0.8$}
    \psfrag{l}[c]{$1.5$}
    \psfrag{m}[c]{$C_P$}
    }
    \caption{Mean pressure coefficient $C_P$ fields from reference $k-\omega$ SST data and 2DVar assimilation.} 
    \label{fig:komega_pressure}
\end{figure}

Additionally, the mean pressure coefficient $C_P$ is presented in Fig. \ref{fig:komega_pressure}, which is calculated as follows:
\begin{align}
C_P = \frac{p - p_{\infty}}{\frac{1}{2}\rho U_{\infty}^2},
\end{align}
where $p_{\infty}$ denotes the freestream pressure, and $\rho$ is the fluid density. The pressure contours exhibit good agreement between the reference and assimilated fields. To further evaluate the accuracy, we compute the discrepancy between the baseline/assimilated pressure fields and the reference pressure field. This discrepancy is calculated as:
\begin{align}
Error_b = \frac{1}{N_c} \sum_{i=1}^{N_c} |C^i_{P_R} - C^i_{P_B}|, \\
Error_a = \frac{1}{N_c} \sum_{i=1}^{N_c} |C^i_{P_R} - C^i_{P_A}|,
\end{align}
where subscripts $b$ and $a$ denote errors relative to the baseline and assimilated fields, respectively. Here, $N_c$ represents the total number of cells in the computational domain, and $i$ corresponds to the $i^{th}$ cell. Additionally, $C_{P_R}$, $C_{P_B}$, and $C_{P_A}$ are the pressure coefficients of the reference, baseline, and assimilated fields, respectively. These values are tabulated in Tab. \ref{tab:komega_pressure}. It is evident that there is a significant improvement in the assimilated mean pressure (by $50\%$) compared to the baseline. 

\section{2DVar assimilation of experimental data}
\label{section:2DVar_assimilation_of_experimental_data}

We showed 2DVar assimilation using synthetic reference data in Sec. \ref{section:assimilation_of_synthetic_data} which is divergence-free for two velocity components. We proceed to perform 2DVar using experimental reference data that do not satisfy 2D continuity computed using Eq. \ref{eq:2D_divergence} as tabulated in Tab. \ref{tab:divergence}. The baseline model used is the SA RANS turbulence model. We use a mesh with a single cell in the spanwise direction, essentially treating the problem as 2D. The experimental reference data are limited to a much smaller domain owing to the reduced field of view obtained through PIV. The observation operator $\mathcal{Q}$ in Eq. \ref{equation:mom_source_obj_func} projects the mean velocity data from the CFD domain within this limited field of view onto the experimental grid, and as such the objective function is calculated only within this region. We examine the mean velocity field and the control variable, i.e. the corrective forcing $f_{c_i}$ and compare it to the Reynolds forcing $f_{R_i}$.

\begin{figure}[h!]
     \centering
         \psfragfig[width=0.8\textwidth]{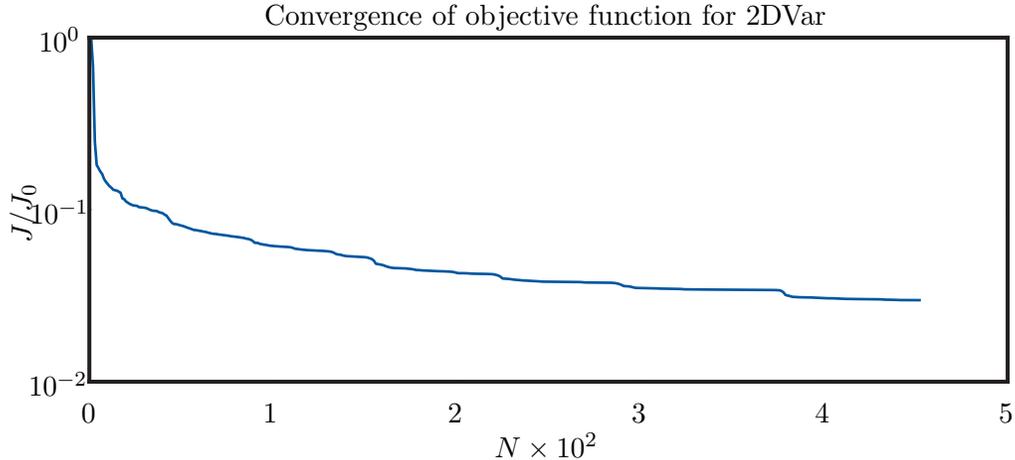}{%
    \psfrag{x}[c]{$10^{-1}$}
    \psfrag{w}[c]{$10^{-2}$}
    \psfrag{y}[c]{$10^{0}$}
    \psfrag{l}[c]{$0$}
    \psfrag{m}[c]{$1$}
    \psfrag{n}[c]{$2$}
    \psfrag{o}[c]{$3$}
    \psfrag{p}[c]{$4$}
    \psfrag{q}[c]{$5$}
    \psfrag{E}[c]{$N \times 10^2$}
    \psfrag{D}[c]{$J/J_0$}
        \psfrag{A}[c]{Convergence of objective function for 2DVar}
    }
    \caption{Convergence of objective function $J$ scaled by its value at the first optimization iteration $J_0$ using 2DVar for experimental reference data. The number of optimization iterations is given by $N$.} 
    \label{fig:doug_2D_convergence}
\end{figure} 

\begin{figure}[h!]
     \centering
         \psfragfig[width=1\textwidth]{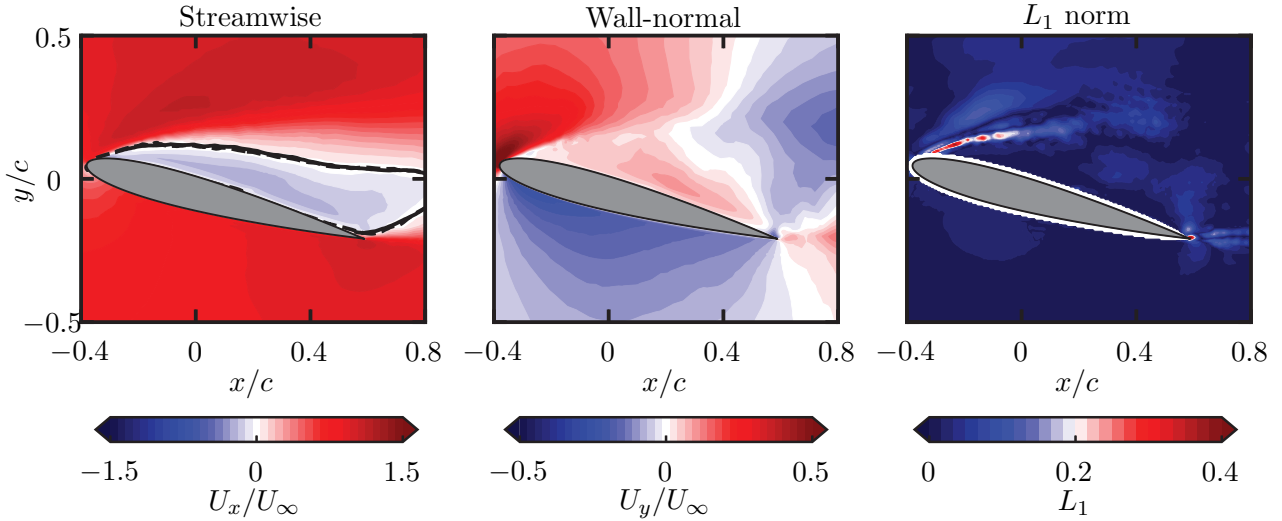}{%
    \psfrag{A}[c]{Streamwise}
    \psfrag{B}[c]{Wall-normal}
    \psfrag{D}[c]{$y/c$}
    \psfrag{E}[c]{$x/c$}
    \psfrag{f}[c]{$-1.5$}
    \psfrag{g}[c]{$-0.4$}
    \psfrag{h}[c]{$-0.5$}
    \psfrag{i}[c]{$0$}
    \psfrag{j}[c]{$0.5$}
    \psfrag{k}[c]{$0.4$}
    \psfrag{o}[c]{$0.8$}
    \psfrag{x}[c]{$0.2$}
    \psfrag{l}[c]{$1.5$}
    \psfrag{m}[c]{$U_x/U_{\infty}$}
    \psfrag{n}[c]{$U_y/U_{\infty}$}
    \psfrag{p}[c]{$L_1$}
    \psfrag{z}[c]{$1$}
        \psfrag{C}[c]{$L_1$ norm}
    }
    \caption{From left to right: 2DVar-assimilated mean streamwise velocity, mean wall-normal velocity fields, and the $L_1$ norm comparing the assimilated mean velocity fields with the experimental data. The $U_x = 0$ contour is depicted as a solid black line for the experimental reference data and as a dashed black line for the data-assimilated model.} 
    \label{fig:doug2D_assm_comp}
\end{figure}

The convergence of the scaled objective function is shown in Fig. \ref{fig:doug_2D_convergence}. The convergence is rapid in the initial iterations, followed by a more gradual change as the optimization progresses. This behavior is expected, as the discrepancy is large at the beginning, prompting the optimizer to apply significant forcing to reduce the discrepancy quickly, with subsequent iterations yielding smaller, more gradual improvements. The scaled objective function converges by over an order of magnitude, indicating successful assimilation of the mean velocity field. Convergence is achieved in just under 500 optimization iterations. The assimilation is performed using $12$ cores of dual $2.0$ GHz Intel Xeon Gold $6138$ processors, resulting in $11.89$ GB of memory usage. Given the parallel computing capabilities of DAFoam on HPC resources, the overall computational cost remains manageable. The assimilated mean streamwise and wall-normal velocity fields are presented in Fig. \ref{fig:doug2D_assm_comp}. The discrepancy field is quantified using the $L_1$ norm. The improvement in the mean velocity fields is evident from the reduced discrepancy in the leading-edge shear layer in Fig. \ref{fig:doug2D_assm_comp}, compared to the baseline. This reduction in discrepancy leads to an increase in the recirculation bubble size, as indicated by the improved agreement between the $U_x = 0$ contours of the experimental and data-assimilated fields. 

To quantify the improvement in the prediction of mean velocity by assimilation, we define an integrated discrepancy which is calculated as, 
\begin{align}
\label{eqn:integrated_disc}
    \epsilon = \frac{1}{N_e} \sum_{i=1}^{N_e} \frac{|\mathcal{Q}(U^i_x) - \hat{U}^i_x| + |\mathcal{Q}(U^i_y) - \hat{U}^i_y|}{U_{\infty}},
\end{align}
where $N_e$ is the number of experimental grid points and the superscript \textit{i} is the $i^{th}$ grid point in the experiment. Equation \ref{eqn:integrated_disc} is just an integral of Eq. \ref{eqn:L1NormPaper}. This is used to compute the relative error $RE$ as
\begin{align}
\label{eqn:relative_error}
    RE = \frac{\epsilon_A}{\epsilon_B},
\end{align}
where $\epsilon_A$ and $\epsilon_B$ are the integrated discrepancies of the assimilated and baseline mean velocity fields,  respectively, compared to the reference data. Smaller values of $RE$ show that the assimilation outperforms the baseline. The values for $\epsilon_B$, $\epsilon_A$, and $RE$ are tabulated in Tab. \ref{tab:integrated_disc_2D}. We observe that the integrated discrepancy of the assimilated model is almost an order of magnitude smaller compared to the baseline which is also reflected in the low magnitude of the $RE$. 

\begin{table}[h!]
    \caption{\label{tab:integrated_disc_2D} Integrated discrepancy and relative error for 2DVar}
    \begin{ruledtabular}
    \renewcommand{\arraystretch}{1.1} 
    \begin{tabular}{@{}l c c  c@{}} 
        & $\epsilon_B$ & $\epsilon_A$ & $RE$  \\ 
        \hline
        Magnitude & $0.164$ & $0.019$ & $0.119$ \\  
    \end{tabular}
    \end{ruledtabular}
\end{table}

To evaluate the divergence of the data-assimilated field, we first compute the mean velocity gradients on the computational grid using the Green-Gauss method. These gradients are then interpolated onto the experimental grid following the procedure described in Appendix \ref{appendix:projection_smoothing}, ensuring that conservation properties are preserved as much as possible. The divergence is subsequently calculated using Eq. \ref{eq:divergence} and is found to be $\nabla \cdot U \approx 5 \times 10^{-4}$. This value is two orders of magnitude smaller than the divergence of the experimental mean velocity field, as reported in Tab. \ref{tab:divergence}. However, discounting interpolation errors, the assimilated mean velocity field can still be considered approximately divergence-free. The discrepancy in divergence between the experimental and assimilated mean velocity fields arises from enforcing a strict 2D divergence-free constraint (Eq. \ref{eq:2D_divergence}) while assimilating a mean velocity field that does not strictly satisfy such a 2D constraint.

\begin{figure}[h!]
     \centering
         \psfragfig[width=0.8\textwidth]{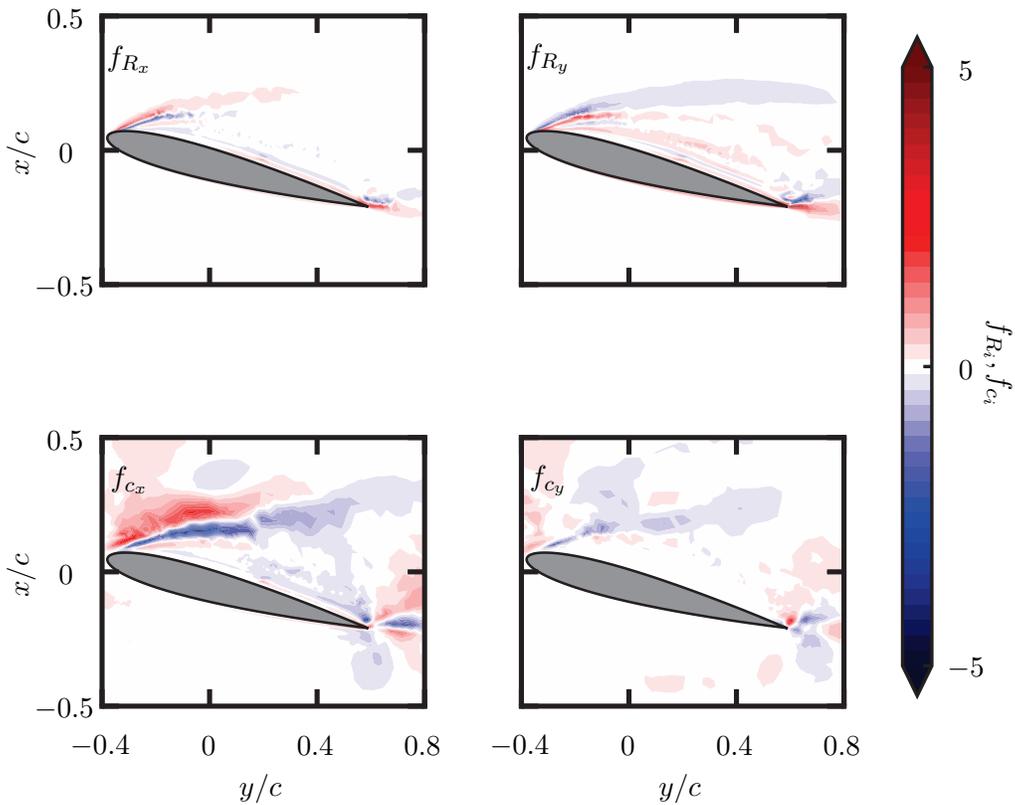}{%
    \psfrag{A}[c]{$f_{R_x}$}
    \psfrag{B}[c]{$f_{R_y}$}
    \psfrag{C}[c]{$f_{c_x}$}
    \psfrag{W}[c]{$f_{c_y}$}
    \psfrag{E}[c]{$y/c$}
    \psfrag{D}[c]{$x/c$}
    \psfrag{g}[c]{$-0.4$}
    \psfrag{h}[c]{$-0.5$}
    \psfrag{i}[c]{$0$}
    \psfrag{j}[c]{$0.5$}
    \psfrag{k}[c]{$0.4$}
    \psfrag{o}[c]{$0.4$}
    \psfrag{n}[c]{$-0.4$}
    \psfrag{p}[c]{$0.8$}
    \psfrag{x}[c]{$-5$}
    \psfrag{z}[c]{$5$}
    \psfrag{m}[c]{$f_{R_i}, f_{c_i}$}
    }
    \caption{Comparison of the streamwise and wall-normal components of the corrective forcing $f_{c_i}$ with the Reynolds forcing $f_{R_i}$ for the 2DVar assimilation using experimental reference data.} 
    \label{fig:doug_compare_optimised_reynolds_forcings_2D}
\end{figure} 

Such a discrepancy in the divergence between the experimental and 2DVar assimilated mean velocity fields directly impacts the corrective forcing field and its interpretation. A comparison between the corrective forcing and Reynolds forcing for the 2DVar assimilated field, using experimental reference data, is presented in Fig. \ref{fig:doug_compare_optimised_reynolds_forcings_2D}. The Reynolds forcing is dominant in the leading- and trailing-edge shear layers. The streamwise component of the corrective forcing exhibits significant spread at the leading edge, with alternating regions of positive and negative magnitude tracing the shear layer. At first glance, it would be reasonable to expect that the forcing in the shear layer entrains freestream flow into the recirculation bubble, leading to its expansion. However, a closer examination reveals that the streamwise corrective forcing aligns with regions in the leading-edge shear layer where the divergence of the experimental velocity field is high, as seen in Fig. \ref{fig:doug_divergence}. This makes it challenging to attribute the effect of the forcing solely to compensating for the lack of physics captured by RANS turbulence models, as it is possible that the forcing is instead addressing the absence of a divergence-free velocity field. 

\begin{table}
    \caption{\label{tab:forcing_exp} Maximum and Minimum Values for Forcing Components}
    \begin{ruledtabular}
    \renewcommand{\arraystretch}{1.3} 
    \begin{tabular}{p{3cm} S S S S} 
        & \multicolumn{2}{l}{\textbf{Corrective Forcing}} & \multicolumn{2}{l}{\textbf{Reynolds Forcing}} \\ 
        & {$f_{c_x}$} & {$f_{c_y}$} & {$f_{R_x}$} & {$f_{R_y}$} \\ 
        \hline
         Minimum & -4.64 & -1.60 & -3.97 & -9.6 \\ 
        Maximum & 3.30 & 2.82 & 3.96 & 5.9 \\   
    \end{tabular}
    \end{ruledtabular}
\end{table}

To quantify these effects, we compare the maximum and minimum values of the corrective forcing with those of the Reynolds forcing in Tab. \ref{tab:forcing_exp}. In general, the corrective forcing values are of comparable magnitude to the Reynolds forcing, except for the minimum value of the wall-normal component. This is in stark contrast to the case of synthetic reference data, where the two forcing terms differ by an order of magnitude. A key factor contributing to this similarity is that the turbulence model cannot account for certain features present in experimental data, such as divergence errors and measurement noise. In this case, the experimental mean velocity field exhibits a significant divergence error, likely due to the inherently 3D nature of the flow, which the 2DVar assimilation is attempting to constrain within a 2D framework. These discrepancies are effectively absorbed into the forcing term. As a result, the corrective forcing not only captures unresolved physical effects in RANS models but also compensates for experimental artifacts. This blending makes it difficult to isolate its true contribution from setup-specific adjustments, posing challenges for symbolic regression, where models may inadvertently reflect experimental-specific features.

\section{3DVar assimilation of experimental data}
\label{section:3DVar_assimilation_of_experimental_data}

We present the results of our novel approach of performing 3DVar assimilation using 2D2C experimental reference data. The key distinction between 3DVar and 2DVar lies in the nature of the constraints enforced. In 3DVar, 3D constraints are enforced through sufficient resolution in the spanwise direction. It will be shown that the enforcement of the 3D continuity constraint (as shown in Eq. \ref{eq:divergence_3D}) allows the development of an out-of-plane component of velocity, confirming the 3D nature of the flow and the inconsistencies in applying a 2D constraint as in 2DVar. 

\begin{figure}[h!]
     \centering
         \psfragfig[width=0.8\textwidth]{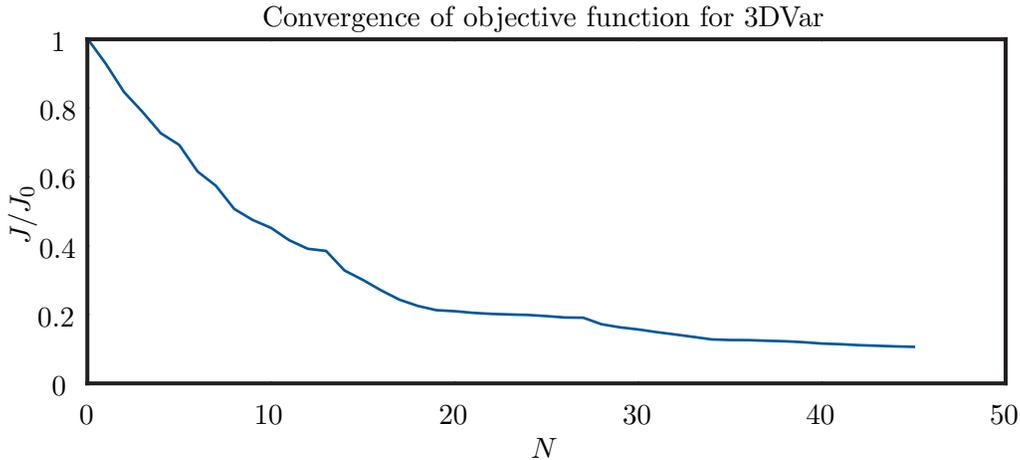}{%
    \psfrag{x}[c]{$0.4$}
    \psfrag{w}[c]{$0.2$}
    \psfrag{y}[c]{$0.6$}
    \psfrag{z}[c]{$0.8$}
    \psfrag{r}[c]{$1$}
    \psfrag{l}[c]{$0$}
    \psfrag{m}[c]{$10$}
    \psfrag{n}[c]{$20$}
    \psfrag{o}[c]{$30$}
    \psfrag{p}[c]{$40$}
    \psfrag{q}[c]{$50$}
    \psfrag{E}[c]{$N$}
    \psfrag{D}[c]{$J/J_0$}
        \psfrag{A}[c]{Convergence of objective function for 3DVar}
    }
    \caption{Convergence of objective function $J$ scaled by its value at the first optimization iteration $J_0$ using 3DVar for experimental reference data. The number of optimization iterations is given by $N$.} 
    \label{fig:doug_3D_convergence}
\end{figure} 

\begin{figure}[h!]
     \centering
         \psfragfig[width=1\textwidth]{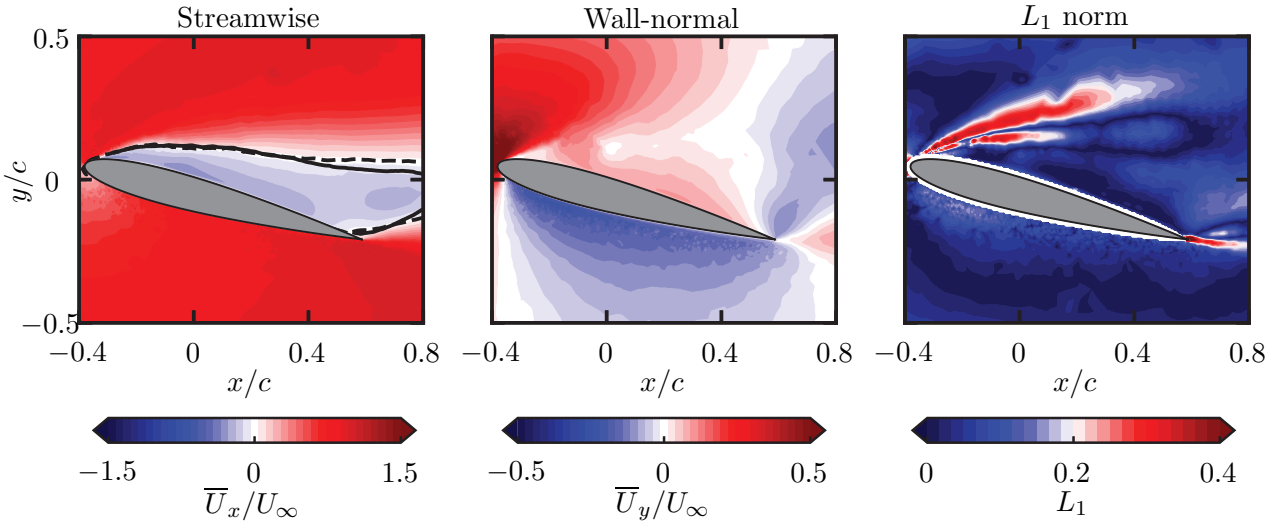}{%
    \psfrag{A}[c]{Streamwise}
    \psfrag{B}[c]{Wall-normal}
    \psfrag{D}[c]{$y/c$}
    \psfrag{E}[c]{$x/c$}
    \psfrag{f}[c]{$-1.5$}
    \psfrag{g}[c]{$-0.4$}
    \psfrag{h}[c]{$-0.5$}
    \psfrag{i}[c]{$0$}
    \psfrag{j}[c]{$0.5$}
    \psfrag{k}[c]{$0.4$}
    \psfrag{o}[c]{$0.8$}
    \psfrag{x}[c]{$0.2$}
    \psfrag{l}[c]{$1.5$}
    \psfrag{m}[c]{$\overline{U}_x/U_{\infty}$}
    \psfrag{n}[c]{$\overline{U}_y/U_{\infty}$}
    \psfrag{p}[c]{$L_1$}
    \psfrag{z}[c]{$1$}
        \psfrag{C}[c]{$L_1$ norm}
    }
    \caption{From left to right: 3DVar-assimilated mean streamwise velocity, mean wall-normal velocity fields, and the $L_1$ norm comparing the assimilated mean velocity fields with the experimental data. The $U_x = 0$ contour is depicted as a solid black line for the experimental reference data and as a dashed black line for the data-assimilated model.} 
    \label{fig:doug3D_assm_comp}
\end{figure}

The convergence of the scaled objective function is presented in Fig. \ref{fig:doug_3D_convergence}. The optimization converges in $45$ iterations with a monotonically decreasing trend. The convergence of the objective function is a lot more gradual in the case of 3DVar as compared to 2DVar (see Fig. \ref{fig:doug_2D_convergence}). The application of 3D constraints, in contrast to 2D constraints, may lead to a less aggressive behavior of the optimizer in adjusting the control variable, as will be discussed in the following section. The assimilation is carried out using 64 cores of dual $2.0$ GHz Intel Xeon Gold $6130$ processors on a high-memory node, resulting in a memory usage of $490$ GB. Although solving adjoint equations using the discrete adjoint method typically demands significant memory, techniques like graph coloring in DAFoam make it feasible to handle this case efficiently. The scaled objective function reduces almost by an order of magnitude, which indicates that the assimilation has taken place successfully. The assimilated mean streamwise and wall-normal velocity fields from 3DVar are presented in Fig. \ref{fig:doug3D_assm_comp}. The data are extracted from the assimilating plane at z=0. The discrepancy along the leading-edge shear layer is reduced, as indicated by the $L_1$ norm, and the 3DVar assimilation correctly predicts the recirculation bubble size. However, the discrepancy field suggests that 3D assimilation fields do not converge as closely to the experimental data as they do in 2DVar. This is reflected in the integrated discrepancy as tabulated in Tab. \ref{tab:integrated_disc_3D}, where $\epsilon_A = 0.061$, which is a little over three times the integrated discrepancy of the 2DVar assimilated field. Some error remains in the leading-edge shear layer that has not been minimized. Despite this, the recirculation region in the assimilated field aligns well with the experimental data.

\begin{table}
    \caption{\label{tab:integrated_disc_3D} Integrated discrepancy and relative error for 3DVar}
    \begin{ruledtabular}
    \renewcommand{\arraystretch}{1.1} 
    \begin{tabular}{@{}l c c  c@{}} 
        & $\epsilon_B$ & $\epsilon_A$ & $RE$  \\ 
        \hline
        Magnitude & $0.164$ & $0.061$ & $0.371$ \\  
    \end{tabular}
    \end{ruledtabular}
\end{table}

One inconsistency in 2DVar is the enforcement of a strict 2D divergence-free constraint, which ensures the assimilated mean velocity field satisfies this condition—even though the experimental reference data does not. In 2DVar, the divergence is computed using only two velocity gradients, as shown in Eq. \ref{eq:2D_divergence}. Here, we extend this analysis to the 3DVar assimilated mean velocity field. Following the same approach as in the 2DVar case, we compute $\partial U_x/\partial x$, $\partial U_y/\partial y$ and $\partial U_z/\partial z$ on the computational mesh using the Green-Gauss method. The gradients along the $z =0$ plane are then extracted and interpolated onto the experimental grid using volume weighting, as described in Appendix \ref{appendix:projection_smoothing}. The divergence of the velocity field is then calculated as
\begin{align}
\label{eq:divergence_3D}
\nabla \cdot U = \frac{1}{N_e} \sum_{N_e} \left(\frac{\partial U_x}{\partial x} + \frac{\partial U_y}{\partial y} + \frac{\partial U_z}{\partial z}\right).  
\end{align}
We compute the integrated spanwise velocity gradient separately to lend weight to the assertion that flow is indeed 3D given by
\begin{align}
\label{eq:dwdz}
\nabla_zU = \frac{1}{N_e} \sum_{N_e} \frac{\partial U_z}{\partial z},  
\end{align}
where $\nabla_zU$ denotes an integrated spanwise velocity gradient of the 3DVar assimilated field. To further assess the consistency between the 3DVar assimilated and experimental fields, we also compute a 2D divergence measure using only $\partial U_x/\partial x$ and $\partial U_y/\partial y$ by applying Eq. \ref{eq:2D_divergence}. This allows us to verify whether the 2D divergence of the 3DVar assimilated mean velocity field matches that of the experiment. The magnitudes of 2D and 3D divergence for the 3DVar assimilated mean velocity field, the divergence of the experimental data, and the integrated spanwise velocity gradient are tabulated in Tab. \ref{tab:divergence_3D}. 

\begin{table}[h!]
    \caption{\label{tab:divergence_3D} Divergence values for 3DVar assimilation}
    \begin{ruledtabular}
    \renewcommand{\arraystretch}{1.1} 
    \begin{tabular}{@{}l c c c c@{}} 
        & Experiment & 2D Divergence & 3D Divergence & $\nabla_zU$ \\ 
        \hline
        Divergence & $-0.0359$ & $-0.0358$ & $-4 \times 10^{-4}$ & $0.0354$ \\  
    \end{tabular}
    \end{ruledtabular}
\end{table}

There are a few important observations to highlight. First, the 2D divergence of the 3DVar assimilated mean velocity field aligns closely with that of the experimental mean velocity field (see Tab. \ref{tab:divergence_3D}). Second, the 3D divergence of the assimilated field is of the same order of magnitude as the 2DVar assimilated field, with the added contribution from the spanwise velocity gradient. By extending the spanwise domain, we have effectively lifted the restrictive 2D continuity constraint, allowing for the development of a spanwise velocity component. The inclusion of this spanwise velocity gradient through the 3D constraints ensures that the 3DVar assimilated field is genuinely divergence-free, while still maintaining agreement in 2D divergence with the experiment. This demonstrates that the flow is inherently 3D, and attempting to assimilate it using 2D constraints is not appropriate.

\begin{figure}[h!]
     \RaggedLeft
         \psfragfig[width=0.9\textwidth]{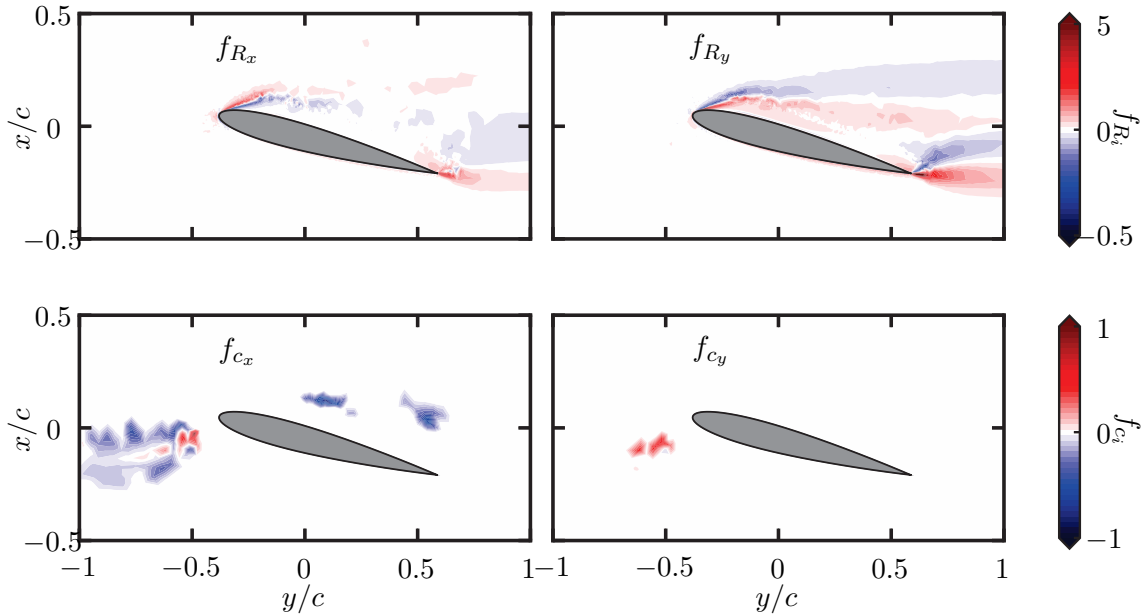}{%
    \psfrag{A}[c]{$f_{R_x}$}
    \psfrag{B}[c]{$f_{R_y}$}
    \psfrag{C}[c]{$f_{c_x}$}
    \psfrag{W}[c]{$f_{c_y}$}
    \psfrag{E}[c]{$y/c$}
    \psfrag{D}[c]{$x/c$}
    \psfrag{g}[c]{$-0.4$}
    \psfrag{h}[c]{$-0.5$}
    \psfrag{i}[c]{$0$}
    \psfrag{j}[c]{$0.5$}
    \psfrag{k}[c]{$0.4$}
    \psfrag{o}[c]{$0.5$}
    \psfrag{n}[c]{$-0.4$}
    \psfrag{p}[c]{$0.8$}
    \psfrag{x}[c]{$-0.5$}
    \psfrag{p}[c]{$1$}
    \psfrag{n}[c]{$-1$}
    \psfrag{z}[c]{$5$}
    \psfrag{y}[c]{$f_{R_i}$}
    \psfrag{w}[c]{$f_{c_i}$}
    }
    \caption{Comparison of the streamwise and wall-normal components of the corrective forcing $f_{c_i}$ with the Reynolds forcing $f_{R_i}$ for the 3DVar assimilation using experimental reference data.} 
    \label{fig:doug_compare_optimised_reynolds_forcings_3D}
\end{figure}

With half the divergence magnitude in 3DVar assimilation coming from $\nabla_zU$ (as computed in Eq. \ref{eq:dwdz}), we turn our attention to the corrective and Reynolds forcing fields. The comparison between these forcing terms is shown in Fig. \ref{fig:doug_compare_optimised_reynolds_forcings_3D}. The streamwise component of the Reynolds forcing remains largely similar to that of 2DVar (Fig. \ref{fig:doug_compare_optimised_reynolds_forcings_2D}), whereas the wall-normal Reynolds forcing exhibits a much broader spread in the leading and trailing edge shear layers. In contrast, the corrective forcing differs significantly from its 2DVar counterpart. Notably, there is some forcing just upstream of the leading edge, which at first glance may appear nonphysical. To interpret this corrective forcing, we consider several sources of uncertainty in the experimental setup, including freestream velocity, airfoil position, and angle of attack. The location and magnitude of the corrective forcing suggest that it is effectively slowing down the velocity at a given angle, potentially accounting for these uncertainties. Additionally, the level of freestream turbulence in the experiment does not match that of the assimilation, meaning this forcing term could also be compensating for that discrepancy. 

\begin{table}[h!]
    \caption{\label{tab:forcing_3d} Maximum and minimum values for forcing components in 3DVar}
    \begin{ruledtabular}
    \renewcommand{\arraystretch}{1.3} 
    \begin{tabular}{p{3cm} S S S S} 
        & \multicolumn{2}{l}{\textbf{Corrective Forcing}} & \multicolumn{2}{l}{\textbf{Reynolds Forcing}} \\ 
        & {$f_{c_x}$} & {$f_{c_y}$} & {$f_{R_x}$} & {$f_{R_y}$} \\ 
        \hline
         Minimum & -1.38 & -1.51 & -7.85 & -6.99 \\ 
        Maximum & 1.23 & 0.85 & 3.99 & 9.10 \\   
    \end{tabular}
    \end{ruledtabular}
\end{table}

An important advantage of 3DVar is its ability to improve the turbulence model’s accuracy while allowing the corrective forcing term to account for errors arising from the experimental setup. This is evident when comparing the magnitudes of the corrective and Reynolds forcing terms, as shown in Tab. \ref{tab:forcing_3d}, similar to the analysis conducted for 2DVar assimilation in Sec. \ref{section:2DVar_assimilation_of_experimental_data}. While the Reynolds forcing magnitude has remained largely unchanged, the corrective forcing has significantly decreased in magnitude. This is an encouraging result, as it reinforces the idea that in 2DVar assimilation, the corrective forcing was primarily compensating for the lack of three-dimensionality in the flow due to the 2D continuity constraint. With the inclusion of a spanwise velocity component in 3DVar assimilation, the baseline RANS model is better able to capture the physics that more accurately represent the experiment.

\section{Reconstructed quantities for 2DVar and 3DVar using experimental reference}
\label{section:reconstructed_quants}
One key advantage of using DA is its ability to derive quantities that are not directly measured during experimental campaigns. While these quantities are not provided as reference data for assimilation, they serve as valuable benchmarks to assess the effectiveness of the assimilation method. We refer to these as reconstructed quantities or reconstructed variables and focus on them in this section. We present a comparison between the reconstructed variables from 2DVar and 3DVar assimilation, further demonstrating that 3DVar outperforms 2DVar, especially when it comes to reconstructed variables. We differentiate between physical reconstructed quantities and modeling reconstructed quantities. Among the physical reconstructed quantities, we focus on examining the pressure field and lift force, as these are of significant physical relevance and can help justify the use of assimilation. In contrast, quantities like eddy viscosity and Reynolds shear stress are reconstructed variables that hold importance within the modeling community.

\begin{figure}[h!]
     \centering
         \psfragfig[width=1\textwidth]{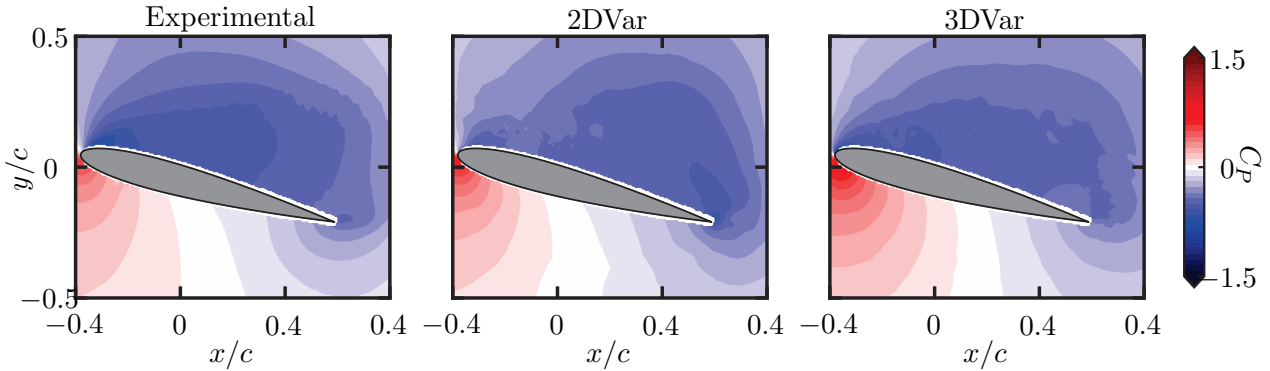}{%
    \psfrag{A}[c]{Experimental}
    \psfrag{B}[c]{2DVar}
    \psfrag{C}[c]{3DVar}
    \psfrag{D}[c]{$y/c$}
    \psfrag{E}[c]{$x/c$}
    \psfrag{f}[c]{$-1.5$}
    \psfrag{g}[c]{$-0.4$}
    \psfrag{h}[c]{$-0.5$}
    \psfrag{i}[c]{$0$}
    \psfrag{r}[c]{$-1$}
    \psfrag{x}[c]{$0.1$}
    \psfrag{y}[c]{$0.2$}
    \psfrag{j}[c]{$0.5$}
    \psfrag{k}[c]{$0.4$}
    \psfrag{o}[c]{$0.4$}
    \psfrag{l}[c]{$1.5$}
    \psfrag{m}[c]{$C_P$}
    }
    \caption{Comparison of mean pressure coefficient $C_P$ fields from experimental data, 2DVar assimilation, and 3DVar assimilation.} 
    \label{fig:doug_2D_3D_pressure}
\end{figure}

Pressure is a quantity that is rarely discussed in most variational DA studies due to the challenges in accurately recovering it when using the Helmholtz decomposition. We present the $C_P$ field computed with a Poisson solver using the mean velocity and Reynolds shear stress fields from the experiment (we refer to this as Poisson pressure hereafter) and compare it with the mean pressure fields obtained from 2DVar and 3DVar assimilation in Fig. \ref{fig:doug_2D_3D_pressure}. The pressure fields from 2DVar and 3DVar assimilation have been interpolated onto the experimental grid using linear interpolation. We must exercise caution when comparing the Poisson pressure and assimilated mean pressure fields.  There are two key caveats to consider: first, the Poisson pressure is derived by applying a Poisson solver to the experimental mean velocity field (see Ref. \cite{de2012instantaneous} for details on this method). Second, the assimilated pressure field includes an irrotational component, $\phi$, as dictated by the Helmholtz decomposition in Eq. \ref{equation:helmholtz}. Therefore, direct comparison between the two pressure fields must take these differences into account.

The experimental pressure field exhibits strong suction at the leading edge on the suction side and a large positive pressure region on the pressure side. The pressure field from 2DVar does not closely resemble the experimental structure, showing a large suction region centered around the trailing edge. On the other hand, the pressure distribution obtained from 3DVar aligns more closely with the experimental field, capturing its key features more accurately. The minor differences may be attributed to the irrotational component in the Helmholtz decomposition.
\begin{figure}[h!]
     \centering
         \psfragfig[width=0.8\textwidth]{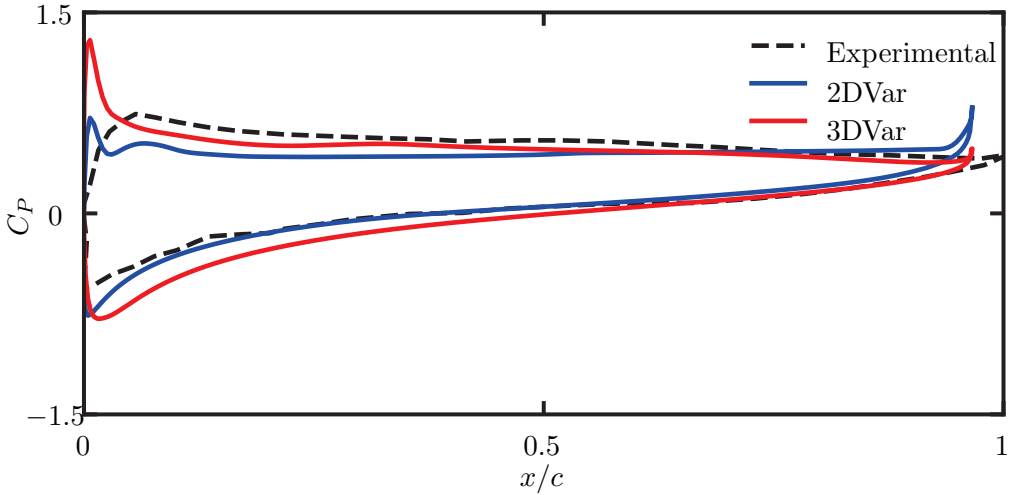}{%
    \psfrag{A}[c]{$C_P$}
    \psfrag{E}[c]{$x/c$}
    \psfrag{P}[l]{Experimental}
    \psfrag{Q}[l]{2DVar}
    \psfrag{R}[l]{3DVar}
    \psfrag{i}[c]{$0$}
    \psfrag{j}[c]{$-1.5$}
    \psfrag{h}[c]{$1.5$}
    \psfrag{k}[c]{$0.5$}
    \psfrag{o}[c]{$1$}
    }
    \caption{Comparison of the surface mean pressure coefficients $C_P$ from experimental data, 2DVar assimilation, and 3DVar assimilation.} 
    \label{fig:cp}
\end{figure}

In addition to examining the pressure fields, we analyze the surface pressure, a key quantity for computing the lift force. A comparison of the surface pressure coefficient between the experiment, 2DVar, and 3DVar assimilation is presented in Fig. \ref{fig:cp}. The suction peak from the experiment is not sharp, likely due to insufficient seeding at the leading edge caused by strong leading edge separation. Since a Poisson solver is used to compute the pressure field, a lack of sufficient velocity resolution affects the pressure calculation. Both 2DVar and 3DVar show a nearly constant surface pressure magnitude on the suction side, a characteristic feature of deep stall, where there is no pressure gradient along the suction side of the airfoil. The 2DVar surface pressure exhibits a small kink near the suction peak and another peak at the trailing edge. This observation aligns with the large suction in the pressure coefficient field seen in Fig. \ref{fig:doug_2D_3D_pressure}. The 2DVar suction-side pressure distribution does not match well with the experiment. In contrast, the 3DVar pressure distribution closely follows the experimental data, with a more pronounced suction peak. 

\begin{table}
    \caption{\label{tab:cl} Comparison of lift coefficient between 2DVar and 3DVar}
    \begin{ruledtabular}
    \renewcommand{\arraystretch}{1.1} 
    \begin{tabular}{@{}l c c c@{}} 
        & Experimental  & 2DVar & 3DVar \\ 
        \hline
        $C_L$  & $0.49$  & $0.43$ & $0.55$ \\  
    \end{tabular}
    \end{ruledtabular}
\end{table}

The experimental setup includes a six-axis load cell, with further details available in Ref. \cite{carter2023low}. The lift force measured by the load cell is reported as the lift coefficient, $C_L$, calculated as

\begin{align} C_L = \frac{L}{\frac{1}{2}\rho U_{\infty}Sc}, \end{align}
where S is the airfoil span. Tab. \ref{tab:cl} compares the $C_L$ values obtained from the load cell with those from 2DVar and 3DVar assimilation. Both 2DVar and 3DVar predictions deviate by approximately $11\%$ from the experimental value in either direction—a notable improvement over the baseline SA model ($C_L = 0.72$). This highlights the advantage of assimilation, as it enables the estimation of lift force without requiring additional measurement devices.

\begin{figure}[h!]
     \centering
         \psfragfig[width=0.8\textwidth]{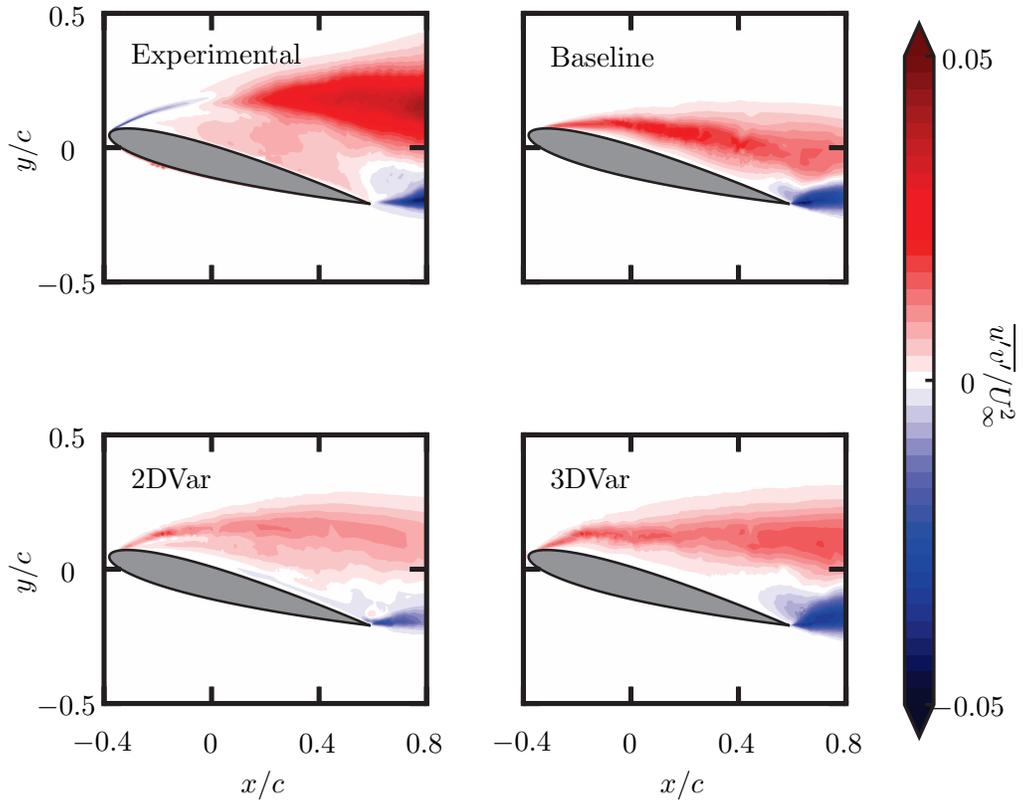}{%
    \psfrag{A}[l]{Experimental}
    \psfrag{B}[l]{Baseline}
    \psfrag{C}[l]{2DVar}
    \psfrag{W}[l]{3DVar}
    \psfrag{D}[c]{$y/c$}
    \psfrag{E}[c]{$x/c$}
    \psfrag{f}[c]{$-1.5$}
    \psfrag{h}[c]{$-0.5$}
    \psfrag{i}[c]{$0$}
    \psfrag{j}[c]{$0.5$}
    \psfrag{l}[c]{$1.5$}
    \psfrag{m}[c]{$\overline{u'v'}/U_{\infty}^2$}
    \psfrag{n}[c]{$-0.4$}
    \psfrag{o}[c]{$0.4$}
    \psfrag{z}[c]{$0.05$}
    \psfrag{x}[c]{$-0.05$}
    \psfrag{p}[c]{$0.8$}
    }
    \caption{Comparison of Reynolds shear stress $\overline{u'v'}/U_\infty^2$ across experimental reference data, baseline SA RANS model, 2DVar, and 3DVar assimilation.} 
    \label{fig:doug2D_rss_comp}
\end{figure} 

We have examined the benefits of 3DVar over 2DVar in reconstructing physically relevant quantities. Now, we shift our focus to reconstructed quantities that hold significance from a modeling perspective. Specifically, we compare the Reynolds shear stress for 2DVar and 3DVar assimilation against the experimental reference in Fig. \ref{fig:doug2D_rss_comp}. Additionally, we include the Reynolds shear stress field from the baseline model for comparison. The baseline model exhibits a positive Reynolds shear stress region along the leading-edge shear layer, which is displaced downward and remains closer to the airfoil surface relative to the experimental reference. This suggests a smaller recirculation bubble, as corroborated by the $U_x = 0$ contour in Fig. \ref{fig:doug2D_bsl_comp}. In contrast, the Reynolds shear stress distributions obtained from both 2DVar and 3DVar assimilation align more closely with the experiment, where the positive Reynolds shear stress region near the leading edge is displaced upward. While both assimilation approaches show improvement, we further investigate the differences observed between 2DVar and 3DVar assimilation.

\begin{figure}[h!]
     \RaggedLeft
         \psfragfig[width=1\textwidth]{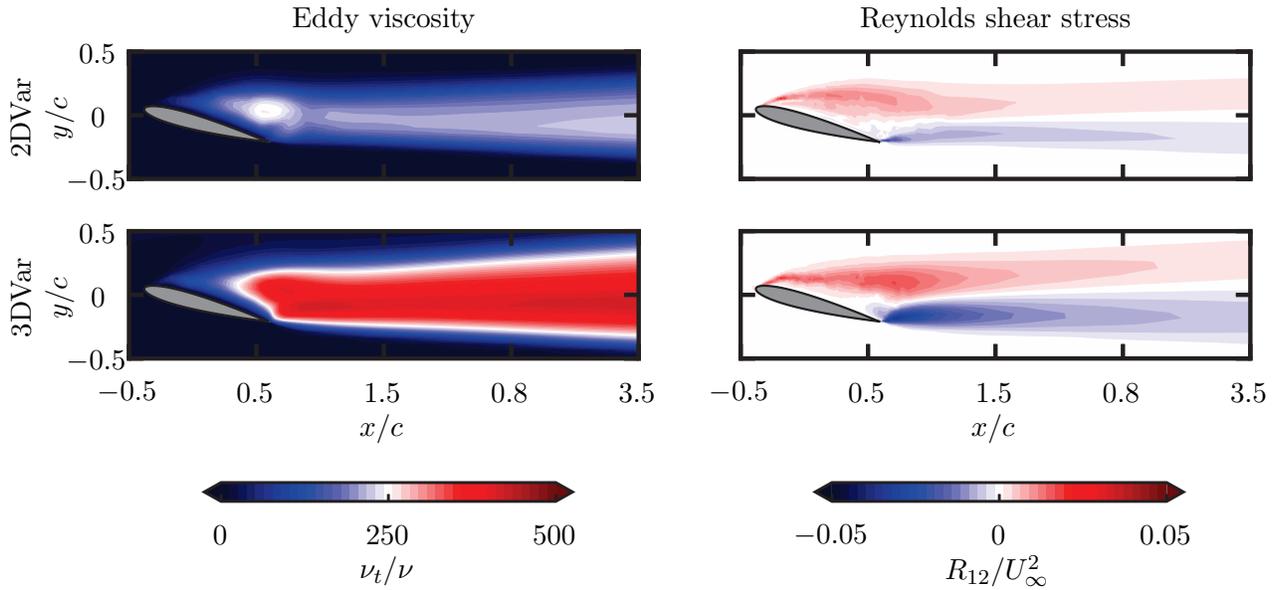}{%
    \psfrag{A}[c]{2DVar}
    \psfrag{B}[c]{3DVar}
    \psfrag{C}[c]{Eddy viscosity}
    \psfrag{W}[c]{Reynolds shear stress}
    \psfrag{D}[c]{$y/c$}
    \psfrag{E}[c]{$x/c$}
    \psfrag{y}[c]{$\nu_t/\nu$}
    \psfrag{w}[c]{$R_{12}/U_{\infty}^2$}
    \psfrag{i}[c]{$0$}
    \psfrag{t}[c]{$250$}
    \psfrag{z}[c]{$500$}
    \psfrag{r}[c]{$-0.05$}
    \psfrag{s}[c]{$0.05$}
    \psfrag{h}[c]{$-0.5$}
    \psfrag{j}[c]{$0.5$}
    \psfrag{o}[c]{$1.5$}
    \psfrag{p}[c]{$2.5$}
    \psfrag{q}[c]{$3.5$}
    \psfrag{x}[c]{$-0.05$}
    \psfrag{p}[c]{$0.8$}
    }
    \caption{Scaled eddy viscosity $\nu_t/\nu$ and Reynolds shear stress comparison between 2DVar and 3DVar assimilation using experimental reference data.} 
    \label{fig:doug_compare_2D_3D_nut_rss}
\end{figure} 

We expand the field of view in the streamwise direction to highlight better the differences in Reynolds shear stress between 2DVar and 3DVar assimilation. Since we employ an SA RANS baseline model, we also analyze the eddy viscosity $\nu_t$, which the SA model utilizes to relate the Reynolds stresses to the mean rate of strain obtained from velocity gradients. The scaled eddy viscosity, $\nu_t/\nu$, and the Reynolds shear stress, $R_{12}/U_\infty^2$, scaled by the freestream velocity, are presented in Fig. \ref{fig:doug_compare_2D_3D_nut_rss}. The majority of the eddy viscosity for 2DVar remains concentrated within the recirculation bubble. In contrast, the eddy viscosity field for 3DVar extends further into the wake of the airfoil. Another key distinction is that the eddy viscosity is significantly stronger for 3DVar than for 2DVar. A stronger eddy viscosity enhances mixing and implies a greater transport of momentum. This trend is also reflected in the distribution of Reynolds shear stress in Fig. \ref{fig:doug_compare_2D_3D_nut_rss}, where the positive component near the leading edge shifts further into the wake, while the negative component at the trailing edge becomes more pronounced. Additionally, the Reynolds shear stress extends further downstream for 3DVar than for 2DVar, indicating a more sustained turbulence effect in the wake and suggesting enhanced momentum transfer over a larger spatial extent. These observations make it clear that the Reynolds shear stress from 3DVar shows a better agreement with the experiment than 2DVar. 

This improvement in the agreement of the Reynolds shear stress from 3DVar and the enhancement of the eddy viscosity in the wake of the airfoil highlight the ability of the baseline turbulence model formulation to capture complex flow physics, such as deep stall at high Reynolds numbers. It provides clear evidence that assimilating inherently 3D flows with 2D constraints distorts the balance of contributions, shifting more of the burden away from the existing turbulence model. Importantly, we have not altered the functional form of the turbulence model. As previously mentioned, the inclusion of a corrective forcing term in the momentum equations, along with 3D constraints, allows the turbulence model to function properly, while the forcing term compensates for deficiencies in the experimental setup. This has significant implications for both the physically relevant quantities (such as surface pressure and lift forces) and those that are critical from a modeling perspective (such as eddy viscosity and Reynolds stress tensor).

\section{Conclusion}
\label{section:conclusion}
We present a novel approach for assimilating 2D2C time-averaged velocity fields obtained from experimental measurements using variational DA with a discrete adjoint method. By enforcing constraints in three dimensions (3DVar), we demonstrate its superiority over conventional variational DA with 2D constraints (2DVar). In both cases, the control variable is introduced as a corrective forcing term in the momentum equations. We apply this method to a deep stall case of a NACA0012 airfoil at a moderate Reynolds number, characterized by strong leading-edge flow separation. Our results show that assimilating an experimental velocity field that is not divergence-free using 2D constraints leads to an increased magnitude of the corrective forcing term, as it compensates for the lack of divergence in the data. This distortion disrupts the balance of the momentum equations and weakens the role of the turbulence model. The true contribution of the corrective forcing term is difficult to isolate, as it accounts for both discrepancies arising from the RANS turbulence model formulation and inconsistencies in the experimental data, including measurement errors, noise, and incomplete information. 

The ambiguity in interpreting the corrective forcing is resolved by implementing 3DVar, where the flow is allowed to develop in the spanwise direction. This enables a more accurate representation of the 3D nature of the flow field while correctly enforcing 3D constraints. We demonstrate that the divergence error in the experimental velocity field is compensated by the spanwise velocity gradient, effectively accounting for missing data in the experiment that caused the divergence error. The resulting corrective forcing field primarily addresses artifacts in the experimental setup, shifting the balance toward the turbulence model through an enhanced Reynolds forcing. This highlights the capability of a baseline RANS model to capture complex flow physics, such as airfoil stall, with greater accuracy than previously assumed, challenging the necessity of directly modifying the momentum equations. We show that 3DVar not only improves the assimilation of mean velocity but also enhances the accuracy of reconstructed quantities, further reinforcing the role of DA in obtaining unmeasured flow properties.

While our 3DVar implementation utilized 2D2C experimental reference data, one of the key outcomes of this study was demonstrating the strengths of 3DVar in assimilating complex fluid dynamics problems at flow conditions of practical relevance. As future work, we plan to extend this approach by incorporating three-component mean velocity fields for assimilation using experimental data. There remains limited clarity on the data requirements for such problems, and few studies have examined how reconstructed quantities are influenced by the amount of data used. Additionally, we aim to perform uncertainty quantification on the obtained corrective forcing, as this remains a topic of significant interest within the DA community.

\section*{Acknowledgments}

The authors acknowledge the use of the IRIDIS High-Performance Computing Facility and its support services at the University of Southampton. We gratefully acknowledge funding from EPSRC (Grant Ref: EP/W009935/1) and the School of Engineering at the University of Southampton for UCP's PhD studentship. The authors declare no conflicts of interest. UCP contributed to the development of DA algorithms, conducted computations, analyzed the data, and drafted multiple versions of the manuscript. BG and SS were responsible for conceptualization, funding acquisition, manuscript editing, and project management.

\appendix
\section{}
\label{appendix:projection_smoothing}
\begin{figure}[!htpb]
    \centering
    \includegraphics[scale=1.5]{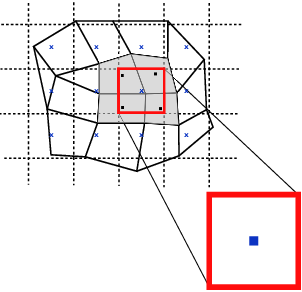}
    \caption{Graphical illustration of the cell volume weighted averaging.  A structured grid (\protect\blackdashed) with experimental data points (\protect\bluecross)  superimposed on an unstructured grid (\protect\blackline) is shown with cell centroids of the CFD mesh (\legendsquare{fill=black}) and cell volume weighted grid (\legendsquare{fill=blue}).} 
    \label{fig:cwva}
\end{figure}

The details of the implementation of projection and smoothing operations are presented here. We consider a field measurement technique such as planar PIV where the output of the processing step is a mean-velocity field defined on a regular cartesian grid of uniform spacing. This is illustrated in Fig. \ref{fig:cwva} as blue crosses. Let us assume that the experimental data co-ordinates are cell centroids of some fictitious cell of uniform dimensions shown as dashed lines. The computational mesh that is used by the DA algorithm is represented as an irregular grid with non-uniform cell size and is superimposed on the regular grid. The 2D representation of this mesh is shown here since each cell is actually a hexahedral element.

Consider one such cartesian grid cell that is outlined in red and enlarged for the sake of explanation. This cell encloses the cell-centroids of four computational mesh cells that have been highlighted in grey. The goal of the algorithm is to find the volume-averaged velocity at the centroid of the cartesian grid cell using the contribution of all the computational mesh cells that lie inside it. Mathematically, this is given by:
\begin{align}
\label{volavg}
    \Bar{u}_g = \dfrac{\sum\limits_{i=1}^N u_iV_i}{\sum\limits_{i=1}^N V_i},
\end{align}
where $\Bar{u}_g$ is the volume-averaged velocity at the grid cell centroid, $N$ is the number of computational mesh cell centroids that lie within a given grid cell (in this example, 4), $u_i$ and $V_i$ are the streamwise velocity and volume of the $i^{th}$ mesh cell, respectively.

The cell-volume weighted averaging procedure can be thought of as the projection operator $\mathcal{Q}$ defined in Eq. \ref{equation:mom_source_obj_func}. The output of this process is the projected computational variables used for calculating the objective function. The objective function is calculated on the regular cartesian grid. As mentioned before, the adjoint solver operates on the computational mesh used in the RANS simulation. This requires a second projection  $\mathcal{\hat{Q}}(.)$ (which we will call smoothing) of the objective function onto the computational mesh. As suggested in \cite{symon2017data}, the objective function at a given Cartesian grid point is distributed equally among all the computational cells that lie inside it.

There is a small modification when performing the projection and smoothing operations for 3D DA. This entails a search in the spanwise direction within the bounds set by the user. Typically, the search is limited to a single layer of cells along the spanwise direction by specifying $z_{min}$ and $z_{max}$ where $\Delta z = z_{max} - z_{min} =  /N_z$, where $N_z$ is the number of points in the spanwise direction. The choice of $N_z$ should then be approximately representative of the laser sheet thickness in a PIV setup. The volume averaging is then restricted to this single layer to avoid the ambiguity of projecting and smoothing more than a single layer of cells. 

\nocite{*}

\bibliography{apssamp}

\providecommand{\noopsort}[1]{}\providecommand{\singleletter}[1]{#1}%
\begin{thebibliography}{47}%
\makeatletter
\providecommand \@ifxundefined [1]{%
 \@ifx{#1\undefined}
}%
\providecommand \@ifnum [1]{%
 \ifnum #1\expandafter \@firstoftwo
 \else \expandafter \@secondoftwo
 \fi
}%
\providecommand \@ifx [1]{%
 \ifx #1\expandafter \@firstoftwo
 \else \expandafter \@secondoftwo
 \fi
}%
\providecommand \natexlab [1]{#1}%
\providecommand \enquote  [1]{``#1''}%
\providecommand \bibnamefont  [1]{#1}%
\providecommand \bibfnamefont [1]{#1}%
\providecommand \citenamefont [1]{#1}%
\providecommand \href@noop [0]{\@secondoftwo}%
\providecommand \href [0]{\begingroup \@sanitize@url \@href}%
\providecommand \@href[1]{\@@startlink{#1}\@@href}%
\providecommand \@@href[1]{\endgroup#1\@@endlink}%
\providecommand \@sanitize@url [0]{\catcode `\\12\catcode `\$12\catcode `\&12\catcode `\#12\catcode `\^12\catcode `\_12\catcode `\%12\relax}%
\providecommand \@@startlink[1]{}%
\providecommand \@@endlink[0]{}%
\providecommand \url  [0]{\begingroup\@sanitize@url \@url }%
\providecommand \@url [1]{\endgroup\@href {#1}{\urlprefix }}%
\providecommand \urlprefix  [0]{URL }%
\providecommand \Eprint [0]{\href }%
\providecommand \doibase [0]{https://doi.org/}%
\providecommand \selectlanguage [0]{\@gobble}%
\providecommand \bibinfo  [0]{\@secondoftwo}%
\providecommand \bibfield  [0]{\@secondoftwo}%
\providecommand \translation [1]{[#1]}%
\providecommand \BibitemOpen [0]{}%
\providecommand \bibitemStop [0]{}%
\providecommand \bibitemNoStop [0]{.\EOS\space}%
\providecommand \EOS [0]{\spacefactor3000\relax}%
\providecommand \BibitemShut  [1]{\csname bibitem#1\endcsname}%
\let\auto@bib@innerbib\@empty
\bibitem [{\citenamefont {Le~Dimet}\ and\ \citenamefont {Talagrand}(1986)}]{le1986variational}%
  \BibitemOpen
  \bibfield  {author} {\bibinfo {author} {\bibfnamefont {F.-X.}\ \bibnamefont {Le~Dimet}}\ and\ \bibinfo {author} {\bibfnamefont {O.}~\bibnamefont {Talagrand}},\ }\bibfield  {title} {\bibinfo {title} {Variational algorithms for analysis and assimilation of meteorological observations: theoretical aspects},\ }\href@noop {} {\bibfield  {journal} {\bibinfo  {journal} {Tellus A: Dyn. Meteorol. Oceanogr.}\ }\textbf {\bibinfo {volume} {38}},\ \bibinfo {pages} {97} (\bibinfo {year} {1986})}\BibitemShut {NoStop}%
\bibitem [{\citenamefont {Foures}\ \emph {et~al.}(2014)\citenamefont {Foures}, \citenamefont {Dovetta}, \citenamefont {Sipp},\ and\ \citenamefont {Schmid}}]{foures2014data}%
  \BibitemOpen
  \bibfield  {author} {\bibinfo {author} {\bibfnamefont {D.~P.~G.}\ \bibnamefont {Foures}}, \bibinfo {author} {\bibfnamefont {N.}~\bibnamefont {Dovetta}}, \bibinfo {author} {\bibfnamefont {D.}~\bibnamefont {Sipp}},\ and\ \bibinfo {author} {\bibfnamefont {P.~J.}\ \bibnamefont {Schmid}},\ }\bibfield  {title} {\bibinfo {title} {A data-assimilation method for {R}eynolds-{A}veraged {N}avier--{S}tokes-driven mean flow reconstruction},\ }\href@noop {} {\bibfield  {journal} {\bibinfo  {journal} {J. Fluid Mech.}\ }\textbf {\bibinfo {volume} {759}},\ \bibinfo {pages} {404} (\bibinfo {year} {2014})}\BibitemShut {NoStop}%
\bibitem [{\citenamefont {Symon}\ \emph {et~al.}(2019)\citenamefont {Symon}, \citenamefont {Sipp},\ and\ \citenamefont {McKeon}}]{symon2019tale}%
  \BibitemOpen
  \bibfield  {author} {\bibinfo {author} {\bibfnamefont {S.}~\bibnamefont {Symon}}, \bibinfo {author} {\bibfnamefont {D.}~\bibnamefont {Sipp}},\ and\ \bibinfo {author} {\bibfnamefont {B.~J.}\ \bibnamefont {McKeon}},\ }\bibfield  {title} {\bibinfo {title} {A tale of two airfoils: resolvent-based modelling of an oscillator versus an amplifier from an experimental mean},\ }\href@noop {} {\bibfield  {journal} {\bibinfo  {journal} {J. Fluid Mech.}\ }\textbf {\bibinfo {volume} {881}},\ \bibinfo {pages} {51} (\bibinfo {year} {2019})}\BibitemShut {NoStop}%
\bibitem [{\citenamefont {Franceschini}\ \emph {et~al.}(2020)\citenamefont {Franceschini}, \citenamefont {Sipp},\ and\ \citenamefont {Marquet}}]{franceschini2020mean}%
  \BibitemOpen
  \bibfield  {author} {\bibinfo {author} {\bibfnamefont {L.}~\bibnamefont {Franceschini}}, \bibinfo {author} {\bibfnamefont {D.}~\bibnamefont {Sipp}},\ and\ \bibinfo {author} {\bibfnamefont {O.}~\bibnamefont {Marquet}},\ }\bibfield  {title} {\bibinfo {title} {Mean-flow data assimilation based on minimal correction of turbulence models: Application to turbulent high {R}eynolds number backward-facing step},\ }\href@noop {} {\bibfield  {journal} {\bibinfo  {journal} {Phys. Rev. Fluids}\ }\textbf {\bibinfo {volume} {5}},\ \bibinfo {pages} {094603} (\bibinfo {year} {2020})}\BibitemShut {NoStop}%
\bibitem [{\citenamefont {Thompson}\ \emph {et~al.}(2024)\citenamefont {Thompson}, \citenamefont {Cadambi~Padmanaban}, \citenamefont {Ganapathisubramani},\ and\ \citenamefont {Symon}}]{thompson2024effect}%
  \BibitemOpen
  \bibfield  {author} {\bibinfo {author} {\bibfnamefont {C.}~\bibnamefont {Thompson}}, \bibinfo {author} {\bibfnamefont {U.}~\bibnamefont {Cadambi~Padmanaban}}, \bibinfo {author} {\bibfnamefont {B.}~\bibnamefont {Ganapathisubramani}},\ and\ \bibinfo {author} {\bibfnamefont {S.}~\bibnamefont {Symon}},\ }\bibfield  {title} {\bibinfo {title} {The effect of variations in experimental and computational fidelity on data assimilation approaches},\ }\href@noop {} {\bibfield  {journal} {\bibinfo  {journal} {Theor. Comput. Fluid. Dyn.}\ }\textbf {\bibinfo {volume} {38}},\ \bibinfo {pages} {431} (\bibinfo {year} {2024})}\BibitemShut {NoStop}%
\bibitem [{\citenamefont {Mons}\ \emph {et~al.}(2024)\citenamefont {Mons}, \citenamefont {Vervynck},\ and\ \citenamefont {Marquet}}]{mons2024data}%
  \BibitemOpen
  \bibfield  {author} {\bibinfo {author} {\bibfnamefont {V.}~\bibnamefont {Mons}}, \bibinfo {author} {\bibfnamefont {A.}~\bibnamefont {Vervynck}},\ and\ \bibinfo {author} {\bibfnamefont {O.}~\bibnamefont {Marquet}},\ }\bibfield  {title} {\bibinfo {title} {Data assimilation and linear analysis with turbulence modelling: application to airfoil stall flows with {P}{I}{V} measurements},\ }\href@noop {} {\bibfield  {journal} {\bibinfo  {journal} {Theor. Comput. Fluid. Dyn.}\ }\textbf {\bibinfo {volume} {38}},\ \bibinfo {pages} {403} (\bibinfo {year} {2024})}\BibitemShut {NoStop}%
\bibitem [{\citenamefont {Pope}(2001)}]{pope2001turbulent}%
  \BibitemOpen
  \bibfield  {author} {\bibinfo {author} {\bibfnamefont {S.~B.}\ \bibnamefont {Pope}},\ }\bibfield  {title} {\bibinfo {title} {Turbulent flows},\ }\href@noop {} {\bibfield  {journal} {\bibinfo  {journal} {Meas. Sci. Technol.}\ }\textbf {\bibinfo {volume} {12}},\ \bibinfo {pages} {2020} (\bibinfo {year} {2001})}\BibitemShut {NoStop}%
\bibitem [{\citenamefont {Adrian}(1991)}]{adrian1991particle}%
  \BibitemOpen
  \bibfield  {author} {\bibinfo {author} {\bibfnamefont {R.~J.}\ \bibnamefont {Adrian}},\ }\bibfield  {title} {\bibinfo {title} {Particle-imaging techniques for experimental fluid mechanics},\ }\href@noop {} {\bibfield  {journal} {\bibinfo  {journal} {Annu. Rev. Fluid. Mech.}\ }\textbf {\bibinfo {volume} {23}},\ \bibinfo {pages} {261} (\bibinfo {year} {1991})}\BibitemShut {NoStop}%
\bibitem [{\citenamefont {Barkley}\ and\ \citenamefont {Henderson}(1996)}]{barkley1996three}%
  \BibitemOpen
  \bibfield  {author} {\bibinfo {author} {\bibfnamefont {D.}~\bibnamefont {Barkley}}\ and\ \bibinfo {author} {\bibfnamefont {R.~D.}\ \bibnamefont {Henderson}},\ }\bibfield  {title} {\bibinfo {title} {Three-dimensional {F}loquet stability analysis of the wake of a circular cylinder},\ }\href@noop {} {\bibfield  {journal} {\bibinfo  {journal} {J. Fluid Mech.}\ }\textbf {\bibinfo {volume} {322}},\ \bibinfo {pages} {215} (\bibinfo {year} {1996})}\BibitemShut {NoStop}%
\bibitem [{\citenamefont {Gupta}\ \emph {et~al.}(2023)\citenamefont {Gupta}, \citenamefont {Zhao}, \citenamefont {Sharma}, \citenamefont {Agrawal}, \citenamefont {Hourigan},\ and\ \citenamefont {Thompson}}]{gupta2023two}%
  \BibitemOpen
  \bibfield  {author} {\bibinfo {author} {\bibfnamefont {S.}~\bibnamefont {Gupta}}, \bibinfo {author} {\bibfnamefont {J.}~\bibnamefont {Zhao}}, \bibinfo {author} {\bibfnamefont {A.}~\bibnamefont {Sharma}}, \bibinfo {author} {\bibfnamefont {A.}~\bibnamefont {Agrawal}}, \bibinfo {author} {\bibfnamefont {K.}~\bibnamefont {Hourigan}},\ and\ \bibinfo {author} {\bibfnamefont {M.~C.}\ \bibnamefont {Thompson}},\ }\bibfield  {title} {\bibinfo {title} {Two-and three-dimensional wake transitions of a {N}{A}{C}{A}0012 airfoil},\ }\href@noop {} {\bibfield  {journal} {\bibinfo  {journal} {J. Fluid Mech.}\ }\textbf {\bibinfo {volume} {954}},\ \bibinfo {pages} {A26} (\bibinfo {year} {2023})}\BibitemShut {NoStop}%
\bibitem [{\citenamefont {Peter}\ and\ \citenamefont {Dwight}(2010)}]{peter2010numerical}%
  \BibitemOpen
  \bibfield  {author} {\bibinfo {author} {\bibfnamefont {J.~E.}\ \bibnamefont {Peter}}\ and\ \bibinfo {author} {\bibfnamefont {R.~P.}\ \bibnamefont {Dwight}},\ }\bibfield  {title} {\bibinfo {title} {Numerical sensitivity analysis for aerodynamic optimization: A survey of approaches},\ }\href@noop {} {\bibfield  {journal} {\bibinfo  {journal} {Comput. Fluids}\ }\textbf {\bibinfo {volume} {39}},\ \bibinfo {pages} {373} (\bibinfo {year} {2010})}\BibitemShut {NoStop}%
\bibitem [{\citenamefont {Kenway}\ \emph {et~al.}(2019)\citenamefont {Kenway}, \citenamefont {Mader}, \citenamefont {He},\ and\ \citenamefont {Martins}}]{kenway2019effective}%
  \BibitemOpen
  \bibfield  {author} {\bibinfo {author} {\bibfnamefont {G.~K.}\ \bibnamefont {Kenway}}, \bibinfo {author} {\bibfnamefont {C.~A.}\ \bibnamefont {Mader}}, \bibinfo {author} {\bibfnamefont {P.}~\bibnamefont {He}},\ and\ \bibinfo {author} {\bibfnamefont {J.~R.}\ \bibnamefont {Martins}},\ }\bibfield  {title} {\bibinfo {title} {Effective adjoint approaches for computational fluid dynamics},\ }\href@noop {} {\bibfield  {journal} {\bibinfo  {journal} {Prog. Aerosp. Sci.}\ }\textbf {\bibinfo {volume} {110}},\ \bibinfo {pages} {100542} (\bibinfo {year} {2019})}\BibitemShut {NoStop}%
\bibitem [{\citenamefont {Symon}\ \emph {et~al.}(2017)\citenamefont {Symon}, \citenamefont {Dovetta}, \citenamefont {McKeon}, \citenamefont {Sipp},\ and\ \citenamefont {Schmid}}]{symon2017data}%
  \BibitemOpen
  \bibfield  {author} {\bibinfo {author} {\bibfnamefont {S.}~\bibnamefont {Symon}}, \bibinfo {author} {\bibfnamefont {N.}~\bibnamefont {Dovetta}}, \bibinfo {author} {\bibfnamefont {B.~J.}\ \bibnamefont {McKeon}}, \bibinfo {author} {\bibfnamefont {D.}~\bibnamefont {Sipp}},\ and\ \bibinfo {author} {\bibfnamefont {P.~J.}\ \bibnamefont {Schmid}},\ }\bibfield  {title} {\bibinfo {title} {Data assimilation of mean velocity from 2{D} {PIV} measurements of flow over an idealized airfoil},\ }\href@noop {} {\bibfield  {journal} {\bibinfo  {journal} {Exp. Fluids}\ }\textbf {\bibinfo {volume} {58}},\ \bibinfo {pages} {1} (\bibinfo {year} {2017})}\BibitemShut {NoStop}%
\bibitem [{\citenamefont {Brenner}\ \emph {et~al.}(2022)\citenamefont {Brenner}, \citenamefont {Piroozmand},\ and\ \citenamefont {Jenny}}]{brenner2022efficient}%
  \BibitemOpen
  \bibfield  {author} {\bibinfo {author} {\bibfnamefont {O.}~\bibnamefont {Brenner}}, \bibinfo {author} {\bibfnamefont {P.}~\bibnamefont {Piroozmand}},\ and\ \bibinfo {author} {\bibfnamefont {P.}~\bibnamefont {Jenny}},\ }\bibfield  {title} {\bibinfo {title} {Efficient assimilation of sparse data into {R}{A}{N}{S}-based turbulent flow simulations using a discrete adjoint method},\ }\href@noop {} {\bibfield  {journal} {\bibinfo  {journal} {J. Comput. Phys.}\ }\textbf {\bibinfo {volume} {471}},\ \bibinfo {pages} {111667} (\bibinfo {year} {2022})}\BibitemShut {NoStop}%
\bibitem [{\citenamefont {Brenner}\ \emph {et~al.}(2024)\citenamefont {Brenner}, \citenamefont {Plogmann}, \citenamefont {Piroozmand},\ and\ \citenamefont {Jenny}}]{brenner2024variational}%
  \BibitemOpen
  \bibfield  {author} {\bibinfo {author} {\bibfnamefont {O.}~\bibnamefont {Brenner}}, \bibinfo {author} {\bibfnamefont {J.}~\bibnamefont {Plogmann}}, \bibinfo {author} {\bibfnamefont {P.}~\bibnamefont {Piroozmand}},\ and\ \bibinfo {author} {\bibfnamefont {P.}~\bibnamefont {Jenny}},\ }\bibfield  {title} {\bibinfo {title} {A variational data assimilation approach for sparse velocity reference data in coarse {R}{A}{N}{S} simulations through a corrective forcing term},\ }\href@noop {} {\bibfield  {journal} {\bibinfo  {journal} {Comput. Methods Appl. Mech. Eng.}\ }\textbf {\bibinfo {volume} {427}},\ \bibinfo {pages} {117026} (\bibinfo {year} {2024})}\BibitemShut {NoStop}%
\bibitem [{\citenamefont {Cato}\ \emph {et~al.}(2023)\citenamefont {Cato}, \citenamefont {Volpiani}, \citenamefont {Mons}, \citenamefont {Marquet},\ and\ \citenamefont {Sipp}}]{cato2023comparison}%
  \BibitemOpen
  \bibfield  {author} {\bibinfo {author} {\bibfnamefont {A.~S.}\ \bibnamefont {Cato}}, \bibinfo {author} {\bibfnamefont {P.~S.}\ \bibnamefont {Volpiani}}, \bibinfo {author} {\bibfnamefont {V.}~\bibnamefont {Mons}}, \bibinfo {author} {\bibfnamefont {O.}~\bibnamefont {Marquet}},\ and\ \bibinfo {author} {\bibfnamefont {D.}~\bibnamefont {Sipp}},\ }\bibfield  {title} {\bibinfo {title} {Comparison of different data-assimilation approaches to augment {RANS} turbulence models},\ }\href@noop {} {\bibfield  {journal} {\bibinfo  {journal} {Comput. Fluids}\ ,\ \bibinfo {pages} {106054}} (\bibinfo {year} {2023})}\BibitemShut {NoStop}%
\bibitem [{\citenamefont {He}\ \emph {et~al.}(2020)\citenamefont {He}, \citenamefont {Mader}, \citenamefont {Martins},\ and\ \citenamefont {Maki}}]{he2020dafoam}%
  \BibitemOpen
  \bibfield  {author} {\bibinfo {author} {\bibfnamefont {P.}~\bibnamefont {He}}, \bibinfo {author} {\bibfnamefont {C.~A.}\ \bibnamefont {Mader}}, \bibinfo {author} {\bibfnamefont {J.~R.}\ \bibnamefont {Martins}},\ and\ \bibinfo {author} {\bibfnamefont {K.~J.}\ \bibnamefont {Maki}},\ }\bibfield  {title} {\bibinfo {title} {{D}{A}{F}oam: An open-source adjoint framework for multidisciplinary design optimization with openfoam},\ }\href@noop {} {\bibfield  {journal} {\bibinfo  {journal} {AIAA J.}\ }\textbf {\bibinfo {volume} {58}},\ \bibinfo {pages} {1304} (\bibinfo {year} {2020})}\BibitemShut {NoStop}%
\bibitem [{\citenamefont {He}\ \emph {et~al.}(2018)\citenamefont {He}, \citenamefont {Mader}, \citenamefont {Martins},\ and\ \citenamefont {Maki}}]{he2018aerodynamic}%
  \BibitemOpen
  \bibfield  {author} {\bibinfo {author} {\bibfnamefont {P.}~\bibnamefont {He}}, \bibinfo {author} {\bibfnamefont {C.~A.}\ \bibnamefont {Mader}}, \bibinfo {author} {\bibfnamefont {J.~R.}\ \bibnamefont {Martins}},\ and\ \bibinfo {author} {\bibfnamefont {K.~J.}\ \bibnamefont {Maki}},\ }\bibfield  {title} {\bibinfo {title} {An aerodynamic design optimization framework using a discrete adjoint approach with {O}pen{FOAM}},\ }\href@noop {} {\bibfield  {journal} {\bibinfo  {journal} {Comput. Fluids}\ }\textbf {\bibinfo {volume} {168}},\ \bibinfo {pages} {285} (\bibinfo {year} {2018})}\BibitemShut {NoStop}%
\bibitem [{\citenamefont {Weller}\ \emph {et~al.}(1998)\citenamefont {Weller}, \citenamefont {Tabor}, \citenamefont {Jasak},\ and\ \citenamefont {Fureby}}]{weller1998tensorial}%
  \BibitemOpen
  \bibfield  {author} {\bibinfo {author} {\bibfnamefont {H.~G.}\ \bibnamefont {Weller}}, \bibinfo {author} {\bibfnamefont {G.}~\bibnamefont {Tabor}}, \bibinfo {author} {\bibfnamefont {H.}~\bibnamefont {Jasak}},\ and\ \bibinfo {author} {\bibfnamefont {C.}~\bibnamefont {Fureby}},\ }\bibfield  {title} {\bibinfo {title} {A tensorial approach to computational continuum mechanics using object-oriented techniques},\ }\href@noop {} {\bibfield  {journal} {\bibinfo  {journal} {Comput. Phys.}\ }\textbf {\bibinfo {volume} {12}},\ \bibinfo {pages} {620} (\bibinfo {year} {1998})}\BibitemShut {NoStop}%
\bibitem [{\citenamefont {Cadambi~Padmanaban}\ \emph {et~al.}(2024)\citenamefont {Cadambi~Padmanaban}, \citenamefont {Ganapathisubramani}, \citenamefont {Vanderwel},\ and\ \citenamefont {Symon}}]{cadambi2024towards}%
  \BibitemOpen
  \bibfield  {author} {\bibinfo {author} {\bibfnamefont {U.}~\bibnamefont {Cadambi~Padmanaban}}, \bibinfo {author} {\bibfnamefont {B.}~\bibnamefont {Ganapathisubramani}}, \bibinfo {author} {\bibfnamefont {C.}~\bibnamefont {Vanderwel}},\ and\ \bibinfo {author} {\bibfnamefont {S.}~\bibnamefont {Symon}},\ }\bibfield  {title} {\bibinfo {title} {Towards passive scalar reconstruction using data assimilation},\ }in\ \href@noop {} {\emph {\bibinfo {booktitle} {Proceedings of the 14th UK Conference on Wind Engineering: University of Southampton}}}\ (\bibinfo {year} {2024})\BibitemShut {NoStop}%
\bibitem [{\citenamefont {Raissi}\ \emph {et~al.}(2019)\citenamefont {Raissi}, \citenamefont {Perdikaris},\ and\ \citenamefont {Karniadakis}}]{raissi2019physics}%
  \BibitemOpen
  \bibfield  {author} {\bibinfo {author} {\bibfnamefont {M.}~\bibnamefont {Raissi}}, \bibinfo {author} {\bibfnamefont {P.}~\bibnamefont {Perdikaris}},\ and\ \bibinfo {author} {\bibfnamefont {G.~E.}\ \bibnamefont {Karniadakis}},\ }\bibfield  {title} {\bibinfo {title} {Physics-informed neural networks: A deep learning framework for solving forward and inverse problems involving nonlinear partial differential equations},\ }\href@noop {} {\bibfield  {journal} {\bibinfo  {journal} {J. Comput. Phys.}\ }\textbf {\bibinfo {volume} {378}},\ \bibinfo {pages} {686} (\bibinfo {year} {2019})}\BibitemShut {NoStop}%
\bibitem [{\citenamefont {Sliwinski}\ and\ \citenamefont {Rigas}(2023)}]{sliwinski2023mean}%
  \BibitemOpen
  \bibfield  {author} {\bibinfo {author} {\bibfnamefont {L.}~\bibnamefont {Sliwinski}}\ and\ \bibinfo {author} {\bibfnamefont {G.}~\bibnamefont {Rigas}},\ }\bibfield  {title} {\bibinfo {title} {Mean flow reconstruction of unsteady flows using physics-informed neural networks},\ }\href@noop {} {\bibfield  {journal} {\bibinfo  {journal} {Data-Centric Eng.}\ }\textbf {\bibinfo {volume} {4}},\ \bibinfo {pages} {e4} (\bibinfo {year} {2023})}\BibitemShut {NoStop}%
\bibitem [{\citenamefont {Patel}\ \emph {et~al.}(2024)\citenamefont {Patel}, \citenamefont {Mons}, \citenamefont {Marquet},\ and\ \citenamefont {Rigas}}]{patel2024turbulence}%
  \BibitemOpen
  \bibfield  {author} {\bibinfo {author} {\bibfnamefont {Y.}~\bibnamefont {Patel}}, \bibinfo {author} {\bibfnamefont {V.}~\bibnamefont {Mons}}, \bibinfo {author} {\bibfnamefont {O.}~\bibnamefont {Marquet}},\ and\ \bibinfo {author} {\bibfnamefont {G.}~\bibnamefont {Rigas}},\ }\bibfield  {title} {\bibinfo {title} {Turbulence model augmented physics-informed neural networks for mean-flow reconstruction},\ }\href@noop {} {\bibfield  {journal} {\bibinfo  {journal} {Phys. Rev. Fluids}\ }\textbf {\bibinfo {volume} {9}},\ \bibinfo {pages} {034605} (\bibinfo {year} {2024})}\BibitemShut {NoStop}%
\bibitem [{\citenamefont {von Saldern}\ \emph {et~al.}(2022)\citenamefont {von Saldern}, \citenamefont {Reumsch{\"u}ssel}, \citenamefont {Kaiser}, \citenamefont {Sieber},\ and\ \citenamefont {Oberleithner}}]{von2022mean}%
  \BibitemOpen
  \bibfield  {author} {\bibinfo {author} {\bibfnamefont {J.~G.}\ \bibnamefont {von Saldern}}, \bibinfo {author} {\bibfnamefont {J.~M.}\ \bibnamefont {Reumsch{\"u}ssel}}, \bibinfo {author} {\bibfnamefont {T.~L.}\ \bibnamefont {Kaiser}}, \bibinfo {author} {\bibfnamefont {M.}~\bibnamefont {Sieber}},\ and\ \bibinfo {author} {\bibfnamefont {K.}~\bibnamefont {Oberleithner}},\ }\bibfield  {title} {\bibinfo {title} {Mean flow data assimilation based on physics-informed neural networks},\ }\href@noop {} {\bibfield  {journal} {\bibinfo  {journal} {Phys. Fluids}\ }\textbf {\bibinfo {volume} {34}} (\bibinfo {year} {2022})}\BibitemShut {NoStop}%
\bibitem [{\citenamefont {Villi{\'e}}\ \emph {et~al.}(2025)\citenamefont {Villi{\'e}}, \citenamefont {Schmitter}, \citenamefont {von Saldern}, \citenamefont {Demange},\ and\ \citenamefont {Oberleithner}}]{villie2025physics}%
  \BibitemOpen
  \bibfield  {author} {\bibinfo {author} {\bibfnamefont {A.}~\bibnamefont {Villi{\'e}}}, \bibinfo {author} {\bibfnamefont {S.}~\bibnamefont {Schmitter}}, \bibinfo {author} {\bibfnamefont {J.~G.}\ \bibnamefont {von Saldern}}, \bibinfo {author} {\bibfnamefont {S.}~\bibnamefont {Demange}},\ and\ \bibinfo {author} {\bibfnamefont {K.}~\bibnamefont {Oberleithner}},\ }\bibfield  {title} {\bibinfo {title} {Physics-informed neural networks for enhancing medical flow magnetic resonance imaging: Artifact correction and mean pressure and reynolds stresses assimilation},\ }\href@noop {} {\bibfield  {journal} {\bibinfo  {journal} {Phys. Fluids}\ }\textbf {\bibinfo {volume} {37}} (\bibinfo {year} {2025})}\BibitemShut {NoStop}%
\bibitem [{\citenamefont {Toma}\ \emph {et~al.}(2025)\citenamefont {Toma}, \citenamefont {Ganapathisubramani},\ and\ \citenamefont {Symon}}]{christian2025review}%
  \BibitemOpen
  \bibfield  {author} {\bibinfo {author} {\bibfnamefont {C.}~\bibnamefont {Toma}}, \bibinfo {author} {\bibfnamefont {B.}~\bibnamefont {Ganapathisubramani}},\ and\ \bibinfo {author} {\bibfnamefont {S.}~\bibnamefont {Symon}},\ }\bibfield  {title} {\bibinfo {title} {Mixed data-source transfer-learning for a turbulence model augmented physics-informed neural network.},\ }\href@noop {} {\bibfield  {journal} {\bibinfo  {journal} {in preparation}\ } (\bibinfo {year} {2025})}\BibitemShut {NoStop}%
\bibitem [{\citenamefont {Mons}\ \emph {et~al.}(2016)\citenamefont {Mons}, \citenamefont {Chassaing}, \citenamefont {Gomez},\ and\ \citenamefont {Sagaut}}]{mons2016reconstruction}%
  \BibitemOpen
  \bibfield  {author} {\bibinfo {author} {\bibfnamefont {V.}~\bibnamefont {Mons}}, \bibinfo {author} {\bibfnamefont {J.-C.}\ \bibnamefont {Chassaing}}, \bibinfo {author} {\bibfnamefont {T.}~\bibnamefont {Gomez}},\ and\ \bibinfo {author} {\bibfnamefont {P.}~\bibnamefont {Sagaut}},\ }\bibfield  {title} {\bibinfo {title} {Reconstruction of unsteady viscous flows using data assimilation schemes},\ }\href@noop {} {\bibfield  {journal} {\bibinfo  {journal} {J. Comput. Phys.}\ }\textbf {\bibinfo {volume} {316}},\ \bibinfo {pages} {255} (\bibinfo {year} {2016})}\BibitemShut {NoStop}%
\bibitem [{\citenamefont {Volpiani}\ \emph {et~al.}(2021)\citenamefont {Volpiani}, \citenamefont {Meyer}, \citenamefont {Franceschini}, \citenamefont {Dandois}, \citenamefont {Renac}, \citenamefont {Martin}, \citenamefont {Marquet},\ and\ \citenamefont {Sipp}}]{volpiani2021machine}%
  \BibitemOpen
  \bibfield  {author} {\bibinfo {author} {\bibfnamefont {P.~S.}\ \bibnamefont {Volpiani}}, \bibinfo {author} {\bibfnamefont {M.}~\bibnamefont {Meyer}}, \bibinfo {author} {\bibfnamefont {L.}~\bibnamefont {Franceschini}}, \bibinfo {author} {\bibfnamefont {J.}~\bibnamefont {Dandois}}, \bibinfo {author} {\bibfnamefont {F.}~\bibnamefont {Renac}}, \bibinfo {author} {\bibfnamefont {E.}~\bibnamefont {Martin}}, \bibinfo {author} {\bibfnamefont {O.}~\bibnamefont {Marquet}},\ and\ \bibinfo {author} {\bibfnamefont {D.}~\bibnamefont {Sipp}},\ }\bibfield  {title} {\bibinfo {title} {Machine learning-augmented turbulence modeling for {R}{A}{N}{S} simulations of massively separated flows},\ }\href@noop {} {\bibfield  {journal} {\bibinfo  {journal} {Phys. Rev. Fluids}\ }\textbf {\bibinfo {volume} {6}},\ \bibinfo {pages} {064607} (\bibinfo {year} {2021})}\BibitemShut {NoStop}%
\bibitem [{\citenamefont {Carter}\ and\ \citenamefont {Ganapathisubramani}(2023)}]{carter2023low}%
  \BibitemOpen
  \bibfield  {author} {\bibinfo {author} {\bibfnamefont {D.~W.}\ \bibnamefont {Carter}}\ and\ \bibinfo {author} {\bibfnamefont {B.}~\bibnamefont {Ganapathisubramani}},\ }\bibfield  {title} {\bibinfo {title} {Low-order modeling and sensor-based prediction of stalled airfoils at moderate reynolds number},\ }\href@noop {} {\bibfield  {journal} {\bibinfo  {journal} {AIAA J.}\ }\textbf {\bibinfo {volume} {61}},\ \bibinfo {pages} {2893} (\bibinfo {year} {2023})}\BibitemShut {NoStop}%
\bibitem [{\citenamefont {Giannakoglou}\ and\ \citenamefont {Papadimitriou}(2008)}]{giannakoglou2008adjoint}%
  \BibitemOpen
  \bibfield  {author} {\bibinfo {author} {\bibfnamefont {K.~C.}\ \bibnamefont {Giannakoglou}}\ and\ \bibinfo {author} {\bibfnamefont {D.~I.}\ \bibnamefont {Papadimitriou}},\ }\bibfield  {title} {\bibinfo {title} {Adjoint methods for shape optimization},\ }\href@noop {} {\bibfield  {journal} {\bibinfo  {journal} {Optim. Comput. Fluid Dyn.}\ ,\ \bibinfo {pages} {79}} (\bibinfo {year} {2008})}\BibitemShut {NoStop}%
\bibitem [{\citenamefont {M.~Sagebaum}(2019)}]{SaAlGauTOMS2019}%
  \BibitemOpen
  \bibfield  {author} {\bibinfo {author} {\bibfnamefont {N.~G.}\ \bibnamefont {M.~Sagebaum}, \bibfnamefont {T.~Albring}},\ }\bibfield  {title} {\bibinfo {title} {High-performance derivative computations using codipack},\ }\href {https://dl.acm.org/doi/abs/10.1145/3356900} {\bibfield  {journal} {\bibinfo  {journal} {ACM Trans. Math. Softw.}\ }\textbf {\bibinfo {volume} {45}} (\bibinfo {year} {2019})}\BibitemShut {NoStop}%
\bibitem [{\citenamefont {Wilson}(1963)}]{wilson1963simplicial}%
  \BibitemOpen
  \bibfield  {author} {\bibinfo {author} {\bibfnamefont {R.~B.}\ \bibnamefont {Wilson}},\ }\bibfield  {title} {\bibinfo {title} {A simplicial algorithm for concave programming},\ }\href@noop {} {\bibfield  {journal} {\bibinfo  {journal} {Ph. D. Dissertation, Graduate School of Bussiness Administration}\ } (\bibinfo {year} {1963})}\BibitemShut {NoStop}%
\bibitem [{\citenamefont {Gill}\ \emph {et~al.}(2005)\citenamefont {Gill}, \citenamefont {Murray},\ and\ \citenamefont {Saunders}}]{gill2005snopt}%
  \BibitemOpen
  \bibfield  {author} {\bibinfo {author} {\bibfnamefont {P.~E.}\ \bibnamefont {Gill}}, \bibinfo {author} {\bibfnamefont {W.}~\bibnamefont {Murray}},\ and\ \bibinfo {author} {\bibfnamefont {M.~A.}\ \bibnamefont {Saunders}},\ }\bibfield  {title} {\bibinfo {title} {{SNOPT}: An {SQP} algorithm for large-scale constrained optimization},\ }\href@noop {} {\bibfield  {journal} {\bibinfo  {journal} {SIAM review}\ }\textbf {\bibinfo {volume} {47}},\ \bibinfo {pages} {99} (\bibinfo {year} {2005})}\BibitemShut {NoStop}%
\bibitem [{\citenamefont {Menter}(1994)}]{menter1994two}%
  \BibitemOpen
  \bibfield  {author} {\bibinfo {author} {\bibfnamefont {F.~R.}\ \bibnamefont {Menter}},\ }\bibfield  {title} {\bibinfo {title} {Two-equation eddy-viscosity turbulence models for engineering applications},\ }\href@noop {} {\bibfield  {journal} {\bibinfo  {journal} {AIAA J.}\ }\textbf {\bibinfo {volume} {32}},\ \bibinfo {pages} {1598} (\bibinfo {year} {1994})}\BibitemShut {NoStop}%
\bibitem [{\citenamefont {Jones}\ and\ \citenamefont {Launder}(1972)}]{jones1972prediction}%
  \BibitemOpen
  \bibfield  {author} {\bibinfo {author} {\bibfnamefont {W.~P.}\ \bibnamefont {Jones}}\ and\ \bibinfo {author} {\bibfnamefont {B.~E.}\ \bibnamefont {Launder}},\ }\bibfield  {title} {\bibinfo {title} {The prediction of laminarization with a two-equation model of turbulence},\ }\href@noop {} {\bibfield  {journal} {\bibinfo  {journal} {Int. J. Heat Mass Transf.}\ }\textbf {\bibinfo {volume} {15}},\ \bibinfo {pages} {301} (\bibinfo {year} {1972})}\BibitemShut {NoStop}%
\bibitem [{\citenamefont {Wilcox}(1988)}]{wilcox1988reassessment}%
  \BibitemOpen
  \bibfield  {author} {\bibinfo {author} {\bibfnamefont {D.~C.}\ \bibnamefont {Wilcox}},\ }\bibfield  {title} {\bibinfo {title} {Reassessment of the scale-determining equation for advanced turbulence models},\ }\href@noop {} {\bibfield  {journal} {\bibinfo  {journal} {AIAA J.}\ }\textbf {\bibinfo {volume} {26}},\ \bibinfo {pages} {1299} (\bibinfo {year} {1988})}\BibitemShut {NoStop}%
\bibitem [{\citenamefont {Patankar}\ and\ \citenamefont {Spalding}(1972)}]{patankar1972calculation}%
  \BibitemOpen
  \bibfield  {author} {\bibinfo {author} {\bibfnamefont {S.}~\bibnamefont {Patankar}}\ and\ \bibinfo {author} {\bibfnamefont {D.}~\bibnamefont {Spalding}},\ }\bibfield  {title} {\bibinfo {title} {A calculation procedure for heat, mass and momentum transfer in three-dimensional parabolic flows},\ }\href@noop {} {\bibfield  {journal} {\bibinfo  {journal} {Int. J. Heat Mass Transf.}\ }\textbf {\bibinfo {volume} {15}},\ \bibinfo {pages} {1787} (\bibinfo {year} {1972})}\BibitemShut {NoStop}%
\bibitem [{\citenamefont {Spalart}\ and\ \citenamefont {Allmaras}(1992)}]{spalart1992one}%
  \BibitemOpen
  \bibfield  {author} {\bibinfo {author} {\bibfnamefont {P.}~\bibnamefont {Spalart}}\ and\ \bibinfo {author} {\bibfnamefont {S.}~\bibnamefont {Allmaras}},\ }\bibfield  {title} {\bibinfo {title} {A one-equation turbulence model for aerodynamic flows},\ }in\ \href@noop {} {\emph {\bibinfo {booktitle} {30th aerospace sciences meeting and exhibit}}}\ (\bibinfo {year} {1992})\ p.\ \bibinfo {pages} {439}\BibitemShut {NoStop}%
\bibitem [{\citenamefont {De~Kat}\ and\ \citenamefont {Van~Oudheusden}(2012)}]{de2012instantaneous}%
  \BibitemOpen
  \bibfield  {author} {\bibinfo {author} {\bibfnamefont {R.}~\bibnamefont {De~Kat}}\ and\ \bibinfo {author} {\bibfnamefont {B.}~\bibnamefont {Van~Oudheusden}},\ }\bibfield  {title} {\bibinfo {title} {Instantaneous planar pressure determination from {P}{I}{V} in turbulent flow},\ }\href@noop {} {\bibfield  {journal} {\bibinfo  {journal} {Exp. Fluids}\ }\textbf {\bibinfo {volume} {52}},\ \bibinfo {pages} {1089} (\bibinfo {year} {2012})}\BibitemShut {NoStop}%
\bibitem [{\citenamefont {Raffel}\ \emph {et~al.}(2018)\citenamefont {Raffel}, \citenamefont {Willert}, \citenamefont {Scarano}, \citenamefont {K{\"a}hler}, \citenamefont {Wereley},\ and\ \citenamefont {Kompenhans}}]{raffel2018particle}%
  \BibitemOpen
  \bibfield  {author} {\bibinfo {author} {\bibfnamefont {M.}~\bibnamefont {Raffel}}, \bibinfo {author} {\bibfnamefont {C.~E.}\ \bibnamefont {Willert}}, \bibinfo {author} {\bibfnamefont {F.}~\bibnamefont {Scarano}}, \bibinfo {author} {\bibfnamefont {C.~J.}\ \bibnamefont {K{\"a}hler}}, \bibinfo {author} {\bibfnamefont {S.~T.}\ \bibnamefont {Wereley}},\ and\ \bibinfo {author} {\bibfnamefont {J.}~\bibnamefont {Kompenhans}},\ }\href@noop {} {\emph {\bibinfo {title} {Particle image velocimetry: a practical guide}}}\ (\bibinfo  {publisher} {Springer},\ \bibinfo {year} {2018})\BibitemShut {NoStop}%
\bibitem [{\citenamefont {Marquet}\ \emph {et~al.}(2022)\citenamefont {Marquet}, \citenamefont {Mons}, \citenamefont {Zauner},\ and\ \citenamefont {Leclaire}}]{marquet2022turbulent}%
  \BibitemOpen
  \bibfield  {author} {\bibinfo {author} {\bibfnamefont {O.}~\bibnamefont {Marquet}}, \bibinfo {author} {\bibfnamefont {V.}~\bibnamefont {Mons}}, \bibinfo {author} {\bibfnamefont {M.}~\bibnamefont {Zauner}},\ and\ \bibinfo {author} {\bibfnamefont {B.}~\bibnamefont {Leclaire}},\ }\bibfield  {title} {\bibinfo {title} {Turbulent mean flow estimation with state observer assimilation of velocity measurements in {R}{A}{N}{S} models},\ }in\ \href@noop {} {\emph {\bibinfo {booktitle} {TSFP 12}}}\ (\bibinfo {year} {2022})\BibitemShut {NoStop}%
\bibitem [{\citenamefont {He}\ \emph {et~al.}(2024)\citenamefont {He}, \citenamefont {Zeng}, \citenamefont {Wang}, \citenamefont {Wen},\ and\ \citenamefont {Liu}}]{he2024four}%
  \BibitemOpen
  \bibfield  {author} {\bibinfo {author} {\bibfnamefont {C.}~\bibnamefont {He}}, \bibinfo {author} {\bibfnamefont {X.}~\bibnamefont {Zeng}}, \bibinfo {author} {\bibfnamefont {P.}~\bibnamefont {Wang}}, \bibinfo {author} {\bibfnamefont {X.}~\bibnamefont {Wen}},\ and\ \bibinfo {author} {\bibfnamefont {Y.}~\bibnamefont {Liu}},\ }\bibfield  {title} {\bibinfo {title} {Four-dimensional variational data assimilation of a turbulent jet for super-temporal-resolution reconstruction},\ }\href@noop {} {\bibfield  {journal} {\bibinfo  {journal} {J. Fluid Mech.}\ }\textbf {\bibinfo {volume} {978}},\ \bibinfo {pages} {A14} (\bibinfo {year} {2024})}\BibitemShut {NoStop}%
\bibitem [{\citenamefont {Gebremedhin}\ \emph {et~al.}(2005)\citenamefont {Gebremedhin}, \citenamefont {Manne},\ and\ \citenamefont {Pothen}}]{gebremedhin2005color}%
  \BibitemOpen
  \bibfield  {author} {\bibinfo {author} {\bibfnamefont {A.~H.}\ \bibnamefont {Gebremedhin}}, \bibinfo {author} {\bibfnamefont {F.}~\bibnamefont {Manne}},\ and\ \bibinfo {author} {\bibfnamefont {A.}~\bibnamefont {Pothen}},\ }\bibfield  {title} {\bibinfo {title} {What color is your {J}acobian? {G}raph coloring for computing derivatives},\ }\href@noop {} {\bibfield  {journal} {\bibinfo  {journal} {SIAM review}\ }\textbf {\bibinfo {volume} {47}},\ \bibinfo {pages} {629} (\bibinfo {year} {2005})}\BibitemShut {NoStop}%
\bibitem [{\citenamefont {de~Silva}\ \emph {et~al.}(2013)\citenamefont {de~Silva}, \citenamefont {Philip},\ and\ \citenamefont {Marusic}}]{de2013minimization}%
  \BibitemOpen
  \bibfield  {author} {\bibinfo {author} {\bibfnamefont {C.~M.}\ \bibnamefont {de~Silva}}, \bibinfo {author} {\bibfnamefont {J.}~\bibnamefont {Philip}},\ and\ \bibinfo {author} {\bibfnamefont {I.}~\bibnamefont {Marusic}},\ }\bibfield  {title} {\bibinfo {title} {Minimization of divergence error in volumetric velocity measurements and implications for turbulence statistics},\ }\href@noop {} {\bibfield  {journal} {\bibinfo  {journal} {Exp. Fluids}\ }\textbf {\bibinfo {volume} {54}},\ \bibinfo {pages} {1} (\bibinfo {year} {2013})}\BibitemShut {NoStop}%
\bibitem [{\citenamefont {Hayase}(2015)}]{hayase2015numerical}%
  \BibitemOpen
  \bibfield  {author} {\bibinfo {author} {\bibfnamefont {T.}~\bibnamefont {Hayase}},\ }\bibfield  {title} {\bibinfo {title} {Numerical simulation of real-world flows},\ }\href@noop {} {\bibfield  {journal} {\bibinfo  {journal} {Fluid Dyn. Res.}\ }\textbf {\bibinfo {volume} {47}},\ \bibinfo {pages} {051201} (\bibinfo {year} {2015})}\BibitemShut {NoStop}%
\bibitem [{\citenamefont {Aulakh}\ \emph {et~al.}(2024)\citenamefont {Aulakh}, \citenamefont {Yang},\ and\ \citenamefont {Maulik}}]{aulakh2024robust}%
  \BibitemOpen
  \bibfield  {author} {\bibinfo {author} {\bibfnamefont {D.~J.~S.}\ \bibnamefont {Aulakh}}, \bibinfo {author} {\bibfnamefont {X.}~\bibnamefont {Yang}},\ and\ \bibinfo {author} {\bibfnamefont {R.}~\bibnamefont {Maulik}},\ }\bibfield  {title} {\bibinfo {title} {Robust experimental data assimilation for the {S}palart-{A}llmaras turbulence model},\ }\href@noop {} {\bibfield  {journal} {\bibinfo  {journal} {Phys. Rev. Fluids}\ }\textbf {\bibinfo {volume} {9}},\ \bibinfo {pages} {084608} (\bibinfo {year} {2024})}\BibitemShut {NoStop}%
\bibitem [{\citenamefont {Zauner}\ \emph {et~al.}(2022)\citenamefont {Zauner}, \citenamefont {Mons}, \citenamefont {Marquet},\ and\ \citenamefont {Leclaire}}]{zauner2021nudging}%
  \BibitemOpen
  \bibfield  {author} {\bibinfo {author} {\bibfnamefont {M.}~\bibnamefont {Zauner}}, \bibinfo {author} {\bibfnamefont {V.}~\bibnamefont {Mons}}, \bibinfo {author} {\bibfnamefont {O.}~\bibnamefont {Marquet}},\ and\ \bibinfo {author} {\bibfnamefont {B.}~\bibnamefont {Leclaire}},\ }\bibfield  {title} {\bibinfo {title} {Nudging-based data assimilation of the turbulent flow around a square cylinder},\ }\href@noop {} {\bibfield  {journal} {\bibinfo  {journal} {J. Fluid Mech.}\ }\textbf {\bibinfo {volume} {937}},\ \bibinfo {pages} {A38} (\bibinfo {year} {2022})}\BibitemShut {NoStop}%
\end{thebibliography}%

\end{document}